\documentclass[journal,10pt,draftclsnofoot,onecolumn]{IEEEtran}
\normalsize
\usepackage{array}
\usepackage{amsmath}
\usepackage{amsmath,graphicx,epsfig,float,times,latexsym,verbatim,mathrsfs,amssymb,textcomp,txfonts,color,setspace}
\usepackage{algorithm}
\usepackage{algorithmic}
\usepackage{url}

\begin{document}

\title{Underwater Cooperative MIMO Communications using Hybrid Acoustic and Magnetic Induction Technique}

\author{Zhangyu Li,
		Soham Desai,
        Vaishnendr D Sudev,
        Pu Wang,
        Jinsong Han,
        and~Zhi~Sun
\thanks{Zhangyu Li, Soham Desai, Vaishnendr D Sudev, and Zhi Sun are with the Department of Electrical Engineering, University at Buffalo, Buffalo, NY 14260 USA (e-mail: zhangyul@buffalo.edu; sohamdhi@buffalo.edu; vaishnen@buffalo.edu; zhisun@buffalo.edu).}
\thanks{Pu Wang is with the Dept. of Computer Science, University of North Carolina at Charlotte, Charlotte, NC, 28223 USA (e-mail: pu.wang@uncc.edu).}
\thanks{Jinsong Han is with the Institute of Cyberspace Research, College of Computer Science and Technology, Zhejiang University, China (e-mail: hanjinsong@zju.edu.cn).}
\thanks{Zhi Sun handles the correspondence of this paper. Phone number: +1 (716) 645-1608. Fax number: +1 (716) 645-3656. E-mail: zhisun@buffalo.edu.}
\thanks{A shorter version of this paper \cite{ourpaper} was presented at the IEEE WCNC 2019.}
}

\maketitle
\thispagestyle{empty}

\begin{abstract}
Future smart ocean applications require long distance and reliable communications to connect underwater sensors/robots with remote surface base stations. It is challenging to achieve such goal due to the harsh and dynamic underwater acoustic channel. While Multiple-Input and Multiple-Output (MIMO) technique can enhance reliability and transmission range, it is difficult to place multiple acoustic transducers on one single underwater device due to the large wavelength. Although the cooperative MIMO technique that let multiple underwater devices form a virtual MIMO system could solve the issue, it was impossible to synchronize the distributed underwater devices due to the extremely large and dynamic propagation delay of acoustic waves. To this end, this paper proposes an underwater cooperative MIMO communication mechanism, which is based on a hybrid acoustic and Magnetic Induction (MI) technique. The inter-node synchronization problem can be perfectly solved by using the MI technique so that the distributed acoustic transducers can cooperatively form narrow beams for long distance underwater communications. The synchronization time and errors are significantly reduced since MI has negligible signal propagation delays. To quantitatively analyze the improvement, the closed-form expressions of the synchronization error, signal-to-noise ratio (SNR), effective communication time, and throughput of the proposed system are rigorously derived. The proposed hybrid system is implemented in a software-defined testbed under the beamforming and space-time coding scheme. Through both numerical analysis and real-world experiments, the paper shows that the proposed hybrid cooperative MIMO mechanism achieves much lower bit error rate and synchronization error than the conventional acoustic systems. 
\end{abstract}


\IEEEpeerreviewmaketitle

\section{Introduction}

The underwater Internet of Things (IoT) have the potential to realize many new smart ocean applications, such as real-time pollution monitoring, deep sea mining, tsunami early warning, tracking underwater animals, and mapping the sea bottom \cite{Akyildiz_softwater_2016, MIT_underwater_backscatter_sigcom_2019, Xu_Underwater_Applications_2014, MariCarmen_UIoT_survay_2012, robot1, robot2, robot3}. Among various underwater IoTs, the Underwater Robotic Swarms (URSs) have attracted more and more attentions. Complex underwater monitoring tasks like recording oceans internal waves in three-dimensional motion are made possible using USRs \cite{int1}. USRs are also able to accomplish complex underwater exploration missions exploiting the collective intelligence that emerges from the local interactions among the robots \cite{int2} \cite{int4} \cite{ursreview}.
This collective intelligence not only requires reliable and real-time communications among the robots within a swarm, but also the long-range, high-throughput, and reliable communication between the robot swarm and the remote surface base station. However, due to the harsh underwater environments, none of the existing wireless networking techniques can simultaneously satisfy the above requirements.

The underwater acoustic Multiple-Input and Multiple-Output (MIMO) system could be used to satisfy the long-range and high-throughput requirements \cite{MIMO3}. However, the size of underwater robots is of the same order of the acoustic wavelength in water (tens of centimeters). As a result, it is impractical to place multiple acoustic transponders in the same underwater robots with enough interspace to guarantee spatial independence (usually more than half wavelength). Moreover, even with MIMO, a single robot still has limited communication range because of its limited onboard power source.

\begin{figure}
\centering
 \includegraphics[width=2.5 in]{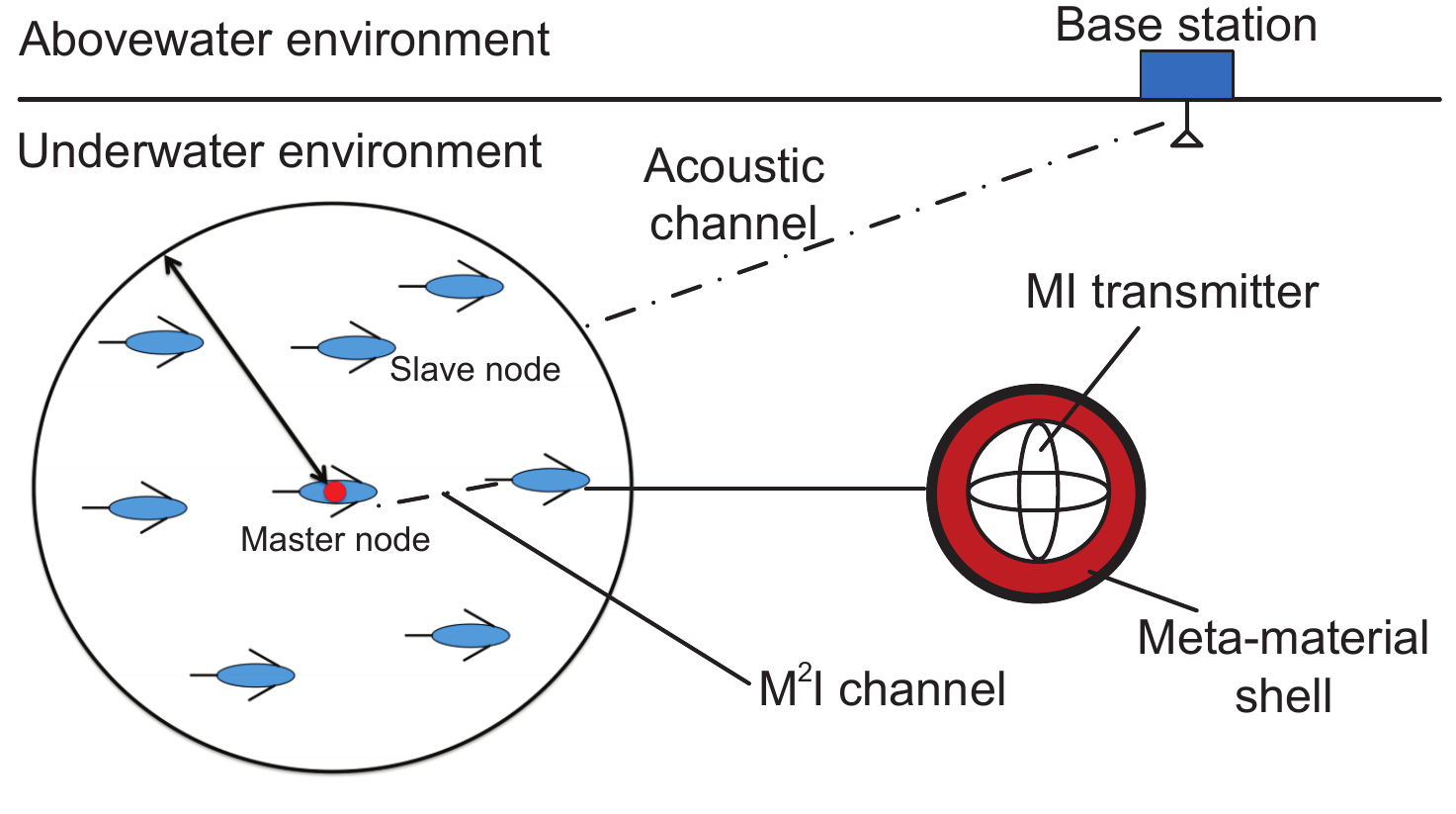}
 \caption{The system overview of the hybrid  M\(^2\)I and acoustic distributed MIMO system. }\label{systemoverview}
 \vspace{-12 pt}
\end{figure}

\begin{figure*}
  \centering
  \includegraphics[width=7in]{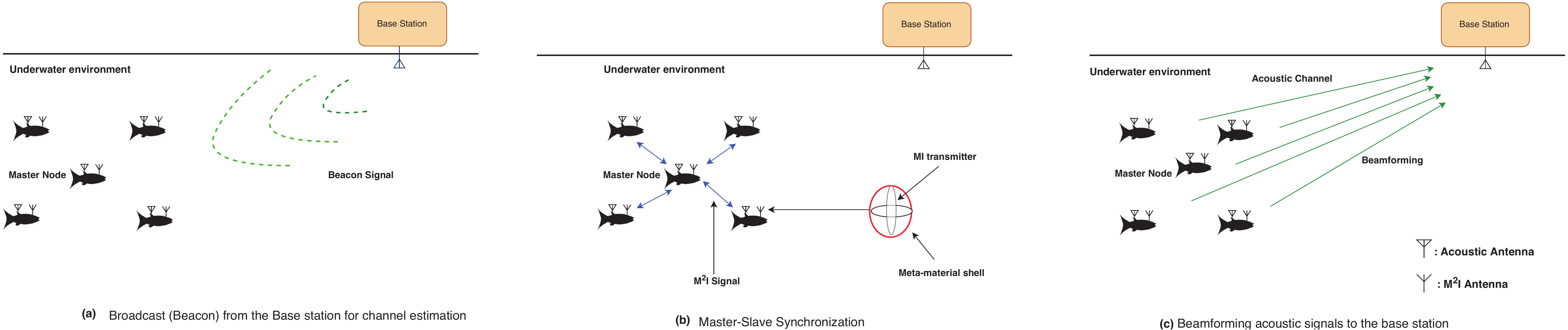}\\
  \caption{The system architecture of the hybrid  M\(^2\)I and acoustic distributed MIMO system.}\label{architecture}
\end{figure*}

Although conventional MIMO is impractical in underwater acoustic communications, the cooperative/distributed MIMO can be utilized. By forming a virtual antenna array using multiple agents, both multiplexing and diversity gains can be achieved. Hence, a distributed MIMO can sufficiently increase the range of wireless links while saving the energy of each transmitting agent. As shown in Fig. \ref{systemoverview}, instead of equipping multiple transmitters on one robot, each robot carries an antenna so that the distributed MIMO is realized by deploying a cluster of robots. As a result, the signal-to-noise ratio (SNR) as well as the channel capacity can be enhanced for the long-range and high-throughput communications. However, unlike conventional MIMO systems, each robot in the cooperative MIMO system has a separate local clock and oscillator. Therefore, the key challenge in realizing the underwater distributed MIMO systems is to synchronize the time and carrier frequency among the transmitters. The main sources of the frequency and time offsets include clock frequency mismatch and oscillator drift, doppler frequency shift, and slow propagation speeds. As a result, acoustic channels cannot satisfy the synchronization requirements since the low propagation speeds and high Doppler sensitivity results in longer synchronization time and larger synchronization errors.

To address the above problem, in this paper, we propose a cooperative MIMO communication mechanism using a hybrid acoustic and magnetic induction (MI) technique \cite{int3}. On the one hand, the MI communication is used for intra-swarm communications. More importantly, MI is also used to synchronize the distributed underwater robots within one swarm and form the virtual MIMO system. On the other hand, once the distributed transducers on each robot is synchronized, the cooperative MIMO acoustic communication is used for long-distance transmission to the remote surface Base Station (BS). MI signals have an underwater propagation speed of $3.33\times10^7$ m/s resulting in a negligible signal propagation delay. Compared with the underwater acoustic waves (propagation speed: $1.4\times10^3$ m/s), the extremely high-speed signal propagation of MI can significantly improve the delay performance while facilitating the synchronization of the distributed transmitters. In addition, the frequency offsets caused by the Doppler shift can also be greatly mitigated while using MI \cite{int3}.
To allow sufficient underwater communication range between robots within one swarm, we also propose using the metamaterial-enhances magnetic induction ($\text{M}^2\text{I}$) in URSs, where the MI transceiver is enclosed in a metamaterial shell (as used in our previous work \cite{meta}) to provide power enhancement.

The major contributions of this paper can be summarized as follows.
\begin{itemize}
  \item We propose the system architecture of the novel hybrid acoustic-MI-based underwater MIMO system, which consists of two modules, namely the $\text{M}^2\text{I}$ assisted synchronization module and the acoustic MIMO module. It is the first time that MI and acoustic techniques are optimally combined in the field of underwater communications.
  \item By considering the synchronization time and clock drift, the time and frequency synchronization errors of both MI and acoustic techniques are rigorously analyzed, which gives a clear comparison between the MI and acoustic techniques in terms of synchronization performance.
  \item The performance of the proposed underwater cooperative MIMO communication system is analytically evaluated under both beamforming and space-time coding schemes. The closed form expressions of the signal-to-noise ratio (SNR), effective communication time, and throughput of the proposed system are rigorously derived, where the system and environmental constrains, such as the clock drift, synchronization errors, and doppler effects, are taken into consideration.
  \item Based on the theoretical analysis, we provide numerical evaluations to prove the significant advantages of the proposed hybrid acoustic-MI MIMO system over the conventional pure acoustic MIMO system.
  \item We implement the proposed hybrid acoustic-MI MIMO system in a software-defined testbed equipped with acoustic hydrophones and $\text{M}^2\text{I}$ transceivers. Two MIMO schemes, i.e., beamforming and space-time coding, are implemented and tested. The system is built in-house and tested in a real water environment (a 8 ft $\times$ 2.5 ft $\times$ 2 ft tank in our lab). Experiment results show that the proposed hybrid MIMO mechanism achieves much lower bit error rate and synchronization error than the conventional acoustic systems.
\end{itemize}

The remainder of the paper is organized as follows. 
The relate work is discussed in section II. Section III gives the system overview. 
The $\text{M}^2\text{I}$-assisted synchronization module is developed in section IV wherein the synchronization protocol and time and frequency synchronization errors are calculated and analyzed. 
Then in section V, the acoustic MIMO module and performance analysis are presented, which is based on the synchronization errors found in the previous section. 
After that, the communication performance of the proposed hybrid system is numerically evaluated in section VI. Details of the software-defined testbed design, system implementations, and experimental results are provided in section VII. Finally, the paper is concluded in Section VIII.

\section{Related Work}
Underwater acoustic channels are band-limited and reverberant, which prevent underwater communication to achieve high throughput and data rate.
Recently, the increasing demand of applications in underwater IoTs draws significant attentions to implement innovative communication techniques in the underwater acoustic networks so that reliable high throughput communications are realized, among which the acoustic MIMO technique is an important paradigm.
In \cite{shallowUAC}, a MIMO acoustic communication system in shadow water environment was proposed and numerically analyzed. 
However, the paper is limited to theoretical concept without real-world implementation. The channel capacity and BER performances can be further improved due to inaccurate channel modeling and lack of coding and protocol design.
In \cite{macmimo, decisionmimo}, the Medium Access Control (MAC) scheme is designed specially for underwater acoustic MIMO system. 
The numerical results showed that the system throughput and the access delay can be significantly improved. However, the synchronization problem of the distributed MIMO is oversimplified and there is still no experimental evaluations. 
In \cite{sonar}, a distributed SONAR MIMO system is designed for target detection and area surveillance. 
However, since the distributed system is connected using wires, it does not have the transmitter synchronization problem as we have in the robotic swarm system. 
The proposed distributed SONAR MIMO system is also limited to theoretical discussion and simulation evaluations.

All the above works focus on theoretical feasibility analysis on the underwater distributed acoustic MIMO communications. However, no practical implementation and real-world experiments are provided.

Real-world implementations of cooperative MIMO system based on electromagnetic (EM) waves have been intensively discussed in terrestrial scenarios \cite{synchronization1}, \cite{synchronization2}, \cite{SDR}, \cite{disbeam}, \cite{jmb}. However, the work on implementations of cooperative MIMO communications in underwater environments is very limited, since it is extremely difficult to synchronize the underwater transmitters due to the low acoustic propagation speed. In \cite{Dmimo}, a distributed underwater MIMO system is implemented, where a MIMO receiver with multiple matched filters at each receiver are designed to counter the multiple carrier frequency offsets (CFO) and time offsets (TO) resulted due to the distributed transmitters. However, the problem of the synchronization amongst the distributed transmitters is not considered at all, which could invalidate the reported promising results in the paper.
In \cite{dopplercoop}, a MIMO receiver is designed and implemented to mitigate the different Doppler affected channel when distributed transmitters are used. However, the synchronization problem of the distributed transmitter is still ignored.

In summary, most of the existing work in this field focus on receiver design assuming very limited or no transceiver synchronization problem. Unless using wire connected distributed transmitters, the synchronization problem using slow acoustic underwater channel cannot be avoided. Therefore, this paper addresses an important issue in the cooperative MIMO underwater communications, where the $\text{M}^2\text{I}$-based intra-swarm communication is used to synchronize the distributed underwater sensors or robots. By combining the $\text{M}^2\text{I}$ and acoustic technique, the long-awaited underwater distributed MIMO system can be truely realized. In addition, it is worth mentioning that our proposed solution synchronizes the transmitters before joint transmission, which can greatly mitigate the problem of multiple CFO and TO at the receiver as well as simplify the receiver design and provide better overall system performance.

\vspace{-6pt}
\section{System Overview}
MIMO system can be used to get both multiplexing and diversity gains.
In this paper, we mostly deal with achieving diversity gains. To this extent, a fixed surface Base Station (BS) is kept as the receiver. The proposed architecture can be extended to obtain multiplexing gains instead when using multiple receivers, i.e. transmitting to a remote robotic swarm.

The aim of the proposed Hybrid acoustic and   M\(^2\)I cooperative MIMO system is establishing long-range and high throughput links between a robot swarm cluster and the surface station. 
An overview of the system is shown in Fig. \ref{systemoverview}. 
In Fig. \ref{systemoverview}, a base station is located on the water surface for the data acquisition. 
The robot swarm is deployed for the detection and exploitation tasks in the underwater environment. 
The proposed hybrid MIMO system can be divided into two modules: the M\(^2\)I-assisted synchronization module and the acoustic MIMO module. 

\vspace{-10pt}
\subsection{ M\(^2\)I-assisted Synchronization Module}

\subsubsection{Introduction to Cooperative Synchronization}

To facilitate synchronization and take control decisions amongst the transmitters, a master node is chosen from the robot swarm and the other nodes work as slave nodes.  
For example, to transmit using either maximum SNR beamforming or space-time coding scheme, all nodes need to agree on which data packets are sent concurrently in one frame. 
This control decision is taken by the master node after exchanging data packets with the slave nodes. 
Moreover, the local hardware clock of the master node is used as the reference to synchronize slave nodes. 
In order to achieve multiplexing gains in beamforming, the channel state information (CSI) of the master node also needs to be delivered to the slave nodes to compute the beamforming codebook \footnote{CSI information is needed since the codebook encoding has to enable concurrent packet decoding at different receivers.}. 
Since MI signals have high propagation speed and large bandwidth in underwater channels, it is used for both these needs. 
Details on the proposed synchronization strategy and results will be discussed in section VII.

\subsubsection{M\(^2\)I-assisted Module}
The M\(^2\)I-assisted synchronization module is mainly based on the tri-dimensional MI transceivers \cite{underwaterMI} enclosed by the metamaterial shells, as shown in  Fig. \ref{systemoverview}.
The MI communication using loop antennas utilizes the magnetic field to carry information.
The tri-dimensional design of the MI transceivers ensures the omni-directionality.
These techniques have inherent advantages for wireless communications in lossy media, especially in underground and underwater environment.
However, underwater robotic swarms require strict size limitation (within 10 cm) of wire-less  transceivers on-board. 
The communication range of MI transceivers is limited since the efficiency becomes low when the size of antennas is much smaller than the MI wavelength (tens of meters). 
Thus, enabling wireless networking with longer communication range is desired in the aforementioned environment and  an artificial meta-material is employed. 
The meta-material shell used here is developed by combining a number of small coils that are uniformly placed on the shell, which is shown in Fig. \ref{MI}.
The meta-material shell can be treated as an array of split resonant ring (SRR) to enhance the magnetic field radiated by the loop antenna.
When the small coils on the shell get resonant at the operating frequency of the loop antennas inside the shell, an negative permeability of metamaterial is achieved, which can enhance the magnetic field  radiated by the loop antenna \cite{M2I}, \cite{metashell}. 
Comparing to MI transceivers without metamaterial shell, the M\(^2\)I transceivers have larger power gain and a 20kHz bandwidth, which can provide low-delay long-range communication links among the robot swarms for the transmitters synchronization and CSI exchange process.

\vspace{-6pt}
\subsection{Acoustic MIMO Module}
Two MIMO implementation schemes are considered in this paper.  
One is maximum SNR beamforming (BF) scheme and second is Space-time block coding (STBC) - Alamouti coding scheme.

\subsubsection{Maximum SNR Beamforming}
In order for the nodes to beamform towards the base station (BS), each distributed antenna need to multiply the data symbols with a complex number corresponding to the inverse of the phase of the channel \footnote{Channel coefficients have to be normalized to respect power constraints.}, hence each source node has to estimate its channel with respect to the BS. 
To achieve this process, the BS broadcasts a beacon signal and a known training sequence as seen in Fig. \ref{architecture}-(a).
Using this signal, the source nodes computed the complex channel at every node. 
Assuming channel reciprocity in the distributed scenario as seen in \cite{realmimo}, the channel from each transmitter to the BS is found. 
Theoretical deductions and implementation details are discussed in section V and VII respectively.

\subsubsection{Space-time Coding}

Space-time block coding uses both spatial and temporal diversity to enable signal gains. 
Just like BF, Space-time coding involves the transmission of multiple copies of data but instead of channel compensation at the transmitter, the data stream has to be encoded in blocks prior to transmission. 
Thus, there is no need for the transmitters to calculate the channel. 
Instead, the encoded data blocks are to be distributed amongst the multiple antennas and the data is spaced across time in a way such that the receivers/BSs are able to decode it effectively.
MIMO Alamouti coding is a very popular and elegant scheme for MIMO STBC.
Hence, we use it during the experimental evaluation of the paper discussed in section VII.

\begin{figure}
  \centering
  \includegraphics[width=2.9in]{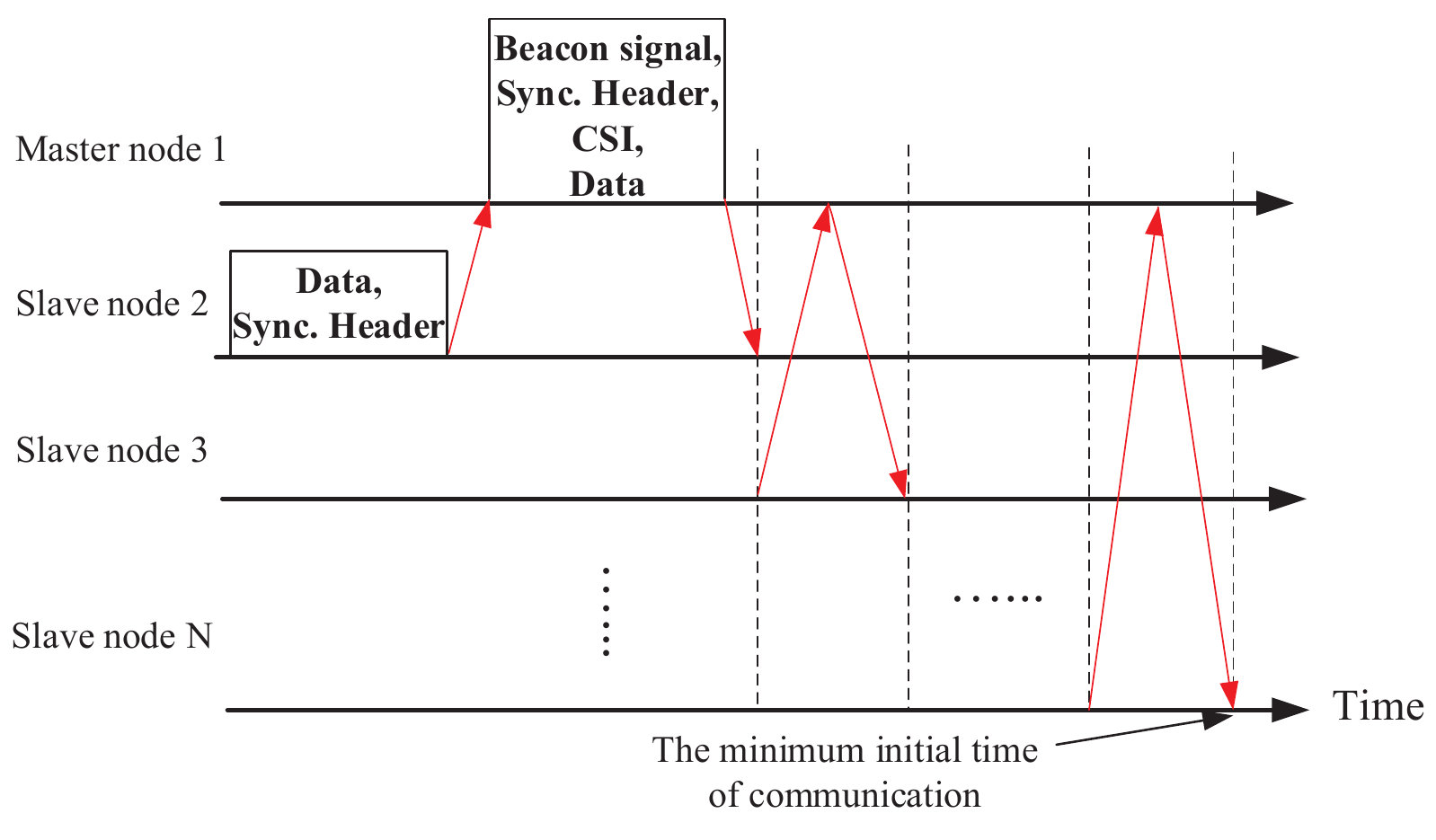}\\
  \caption{The master-slave synchronization of underwater distributed MIMO systems.}\label{protocol}
  \vspace{-9pt}
\end{figure}

\vspace{-6pt}
\section{$\text{M}^2\text{I}$-assisted Synchronization Module}
In this section, we first discuss the proposed strategy for transmitter synchronization. 
Then, we analyze the accuracy for frequency and time synchronization using  M\(^2\)I and pure acoustic based communication.

\vspace{-9pt}
\subsection{Synchronization Protocol}
As shown in Fig. \ref{protocol}, the synchronization strategy works as follows: A master node is predetermined from the transmitting nodes and the other nodes use the master node as the reference. 
Since all the nodes have the same characteristics and  M\(^2\)I antennas used for synchronization have a relatively small time delay, the random assignment of the master will not have an impact on the overall system performance. 
The synchronization starts immediately after the BS broadcasts the initial beacon signal. 
The master node follows the initial BS broadcast with another broadcast to all the slave nodes using $\text{M}\text{I}$ signals. 
This broadcast contains a beacon signal at a known frequency and a synchronization preamble which is used for carrier frequency synchronization at each slave node. After this, the time/phase synchronization happens in a time division duplexing (TDD) order utilizing  M\(^2\)I communication between the master and slave nodes as seen in Fig. \ref{protocol} and based on the method described in \cite{metaextreme}, \cite{coherent}. 
The necessary data and CSI information needed for the generation of beamforming codebooks are also exchanged in this stage. 

Compared to the acoustic channels, the  M\(^2\)I channels between the transmitting nodes have a very tiny delay.
As a result, the synchronization time can be reduced and therefore the synchronization accuracy is significantly increased. 
The numerical analysis is formulated in the following section.

\vspace{-6pt}
\subsection{Analysis of Synchronization Accuracy}
\vspace*{-1.5mm}
First, we analyze the frequency synchronization error. 
Considering the difference of the independent crystal oscillators in the transmitter nodes, the average relative clock drift of the $n$-th node to the master node in the time $\Delta T$ can be presented as
\begin{equation}\label{average}
  \bar{a}_n=\frac{\int_{\Delta T} a_n(t) dt}{\Delta T},
\end{equation}
where $a_n(t)$ is the time-varying drift defined as the ratio of oscillator frequencies, we have
\begin{equation}\label{ratio}
  a_n(t)=\frac{f_{c,n}(t)}{f_{c,1}(t)},
\end{equation}
where $f_{c,n}(t)$ is the oscillator frequency of node $n$, the subscript $n = 2, 3, ..., N$ indicates the slave nodes. 
The subscript $1$ indicates the master node.

For a certain transmitting node, the operating frequency $f_{s,i}(t)$ is proportional to its hardware oscillator, which can be expressed as
\begin{equation}\label{operating}
  f_{s,i}(t)=k \cdot f_{c,i}(t)     \quad \forall  i=1,2, ...N,
\end{equation}
where $k$ is the frequency multiplier. 
The frequency of the beacon signal generated by the master node is $f_{s,1}$. 
After receiving the beacon signal at the slave node $n$, the frequency of beacon signal is estimated according to the local oscillator of the slave node as
\begin{equation}\label{frequencyest}
  \hat{f}_{s,1,n}=\bar{a}_n f_{s,1}+\epsilon_{s,n},
\end{equation}
where $\epsilon_{s,n}$ is the error of the frequency estimation. 
Obviously, the frequency estimated by the slave node $n$,  $\hat{f}_{s,1,n}$ is different from the transmitting frequency $f_{s,1}$ due to the relative clock drift and the estimation error. 
Meanwhile, the slave node is told by the master node that the frequency of the beacon signal is $f_{s,1}$. Therefore, the difference of the frequency can be calculated as
\begin{equation}\label{difference}
  \hat{f}_{s,1,n} - f_{s,1} = (\bar{a}_n -1) f_{s,1} + \epsilon_{s,n}.
\end{equation}
The frequency offset at the slave node $n$ for the frequency synchronization can be determined as
\begin{equation}\label{offset}
  \Delta f_{c,n} = \frac{\hat{f}_{s,1,n} - f_{s,1}}{k} = (\bar{a}_n -1) f_{c,1} + \frac{\epsilon_{s,n}}{k}.
\end{equation}
According to (\ref{offset}), the optimal frequency offset is $(\bar{a}_n -1) f_{c,1}$. 
Due to the estimation error $\epsilon_{s,n}$, the frequency cannot be perfectly synchronized and the frequency synchronization error can be defined as
\begin{equation}\label{frequencyerror}
\epsilon_{n,f}=\frac{\epsilon_{s,n}}{k}.
\end{equation}

Then, we analyze the time error of the synchronization. 
Due to the clock drift caused by the difference of oscillators, the accuracy of both time and frequency synchronization decreases as the drift bound increases. Once the master node generates its time stamp to slave node $n$, for the time synchronization process, the clocks of slave nods begin to relatively drift so that they will not start the beamforming at exactly the same time. 
According to the polling operation shown in Fig. \ref{protocol}, the duration of the time slot for the $n$-th node is calculated by
\begin{equation}\label{drifttime}
  \Delta t_n = \frac{L_{n,1}+L_{1,n}}{B} + \frac{2 d_{1,n}}{v},
\end{equation}
where $L_{n,1}$ and $L_{1,n}$ are the total packet length delivered from the slave node to the master node and that from the master node to the slave node, respectively. 
$B$ is the bandwidth, $d_{1,n}$ is the distance between the master node and the slave node $n$, $v$ is the propagation speed of the signals used for synchronization. 
Obviously, compared to the acoustic channel, the M\(^2\)I channel has larger bandwidth and propagation speed. 
Hence, the duration of each time slot can be reduced if the  M\(^2\)I communication is utilized.

The total drifting time for the slave node $n$ is computed from the timing that the master node generates the time stamp for node $n$ to the earliest beamforming time $t_{BF}$.
Considering the time of the master node as the reference, the time synchronization error of slave node $n$ can be derived according to (\ref{frequencyerror}) and (\ref{drifttime})
\begin{equation}\label{timeerror}
  \epsilon_{n,t} =\epsilon_{n,f} \left(\frac{L_{1,n}}{B} + \frac{d_{1,n}}{v} + \sum^{N}_{i=n+1} \Delta t_i\right).
\end{equation}

\begin{figure}
  \centering
  \includegraphics[width=2.6in]{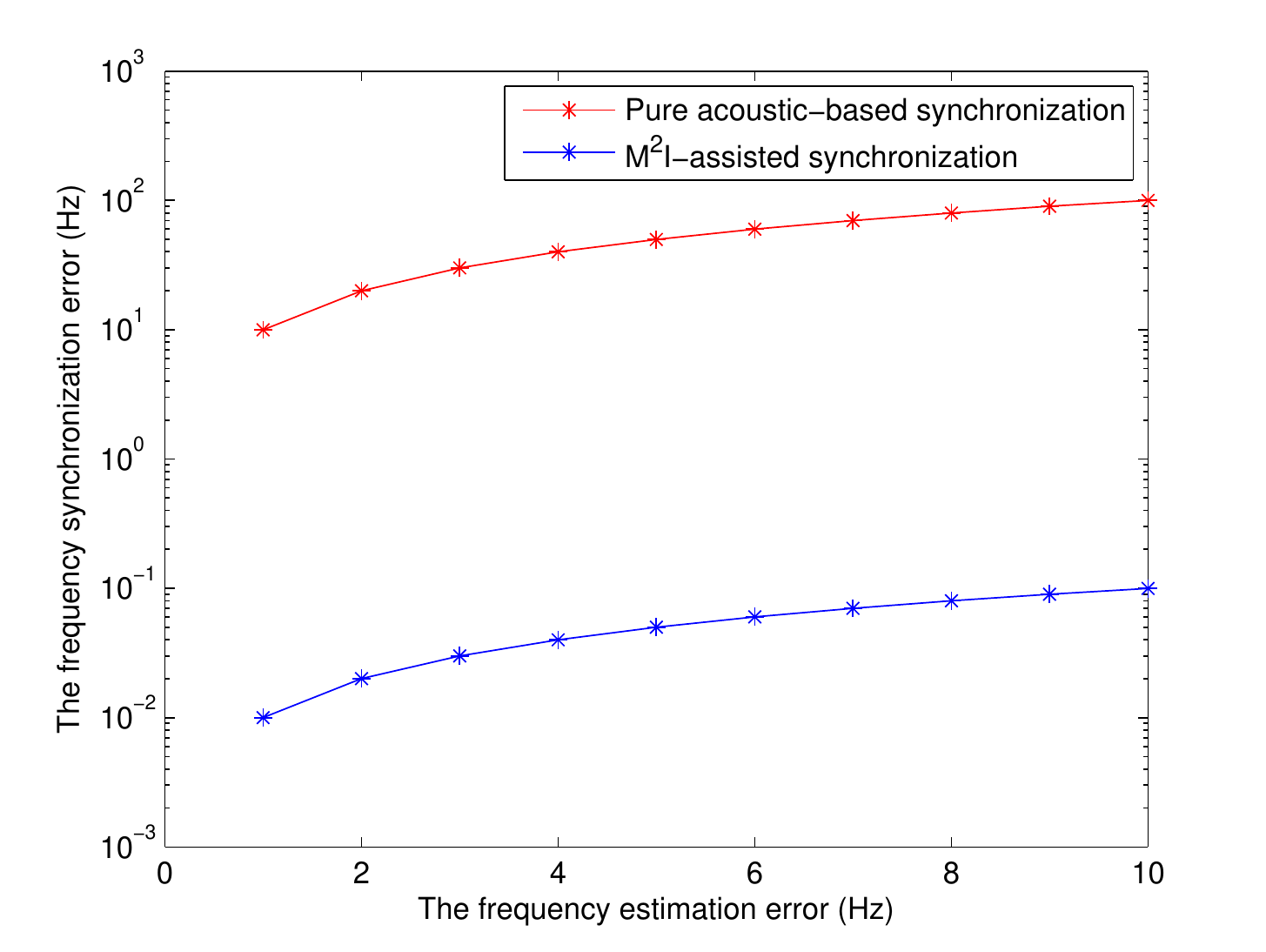}\\
  \caption{The frequency synchronization error of  M\(^2\)I-assisted and pure acoustic based communication.}\label{frequencyerr}
\end{figure}
\begin{figure}
  \centering
  \includegraphics[width=2.6in]{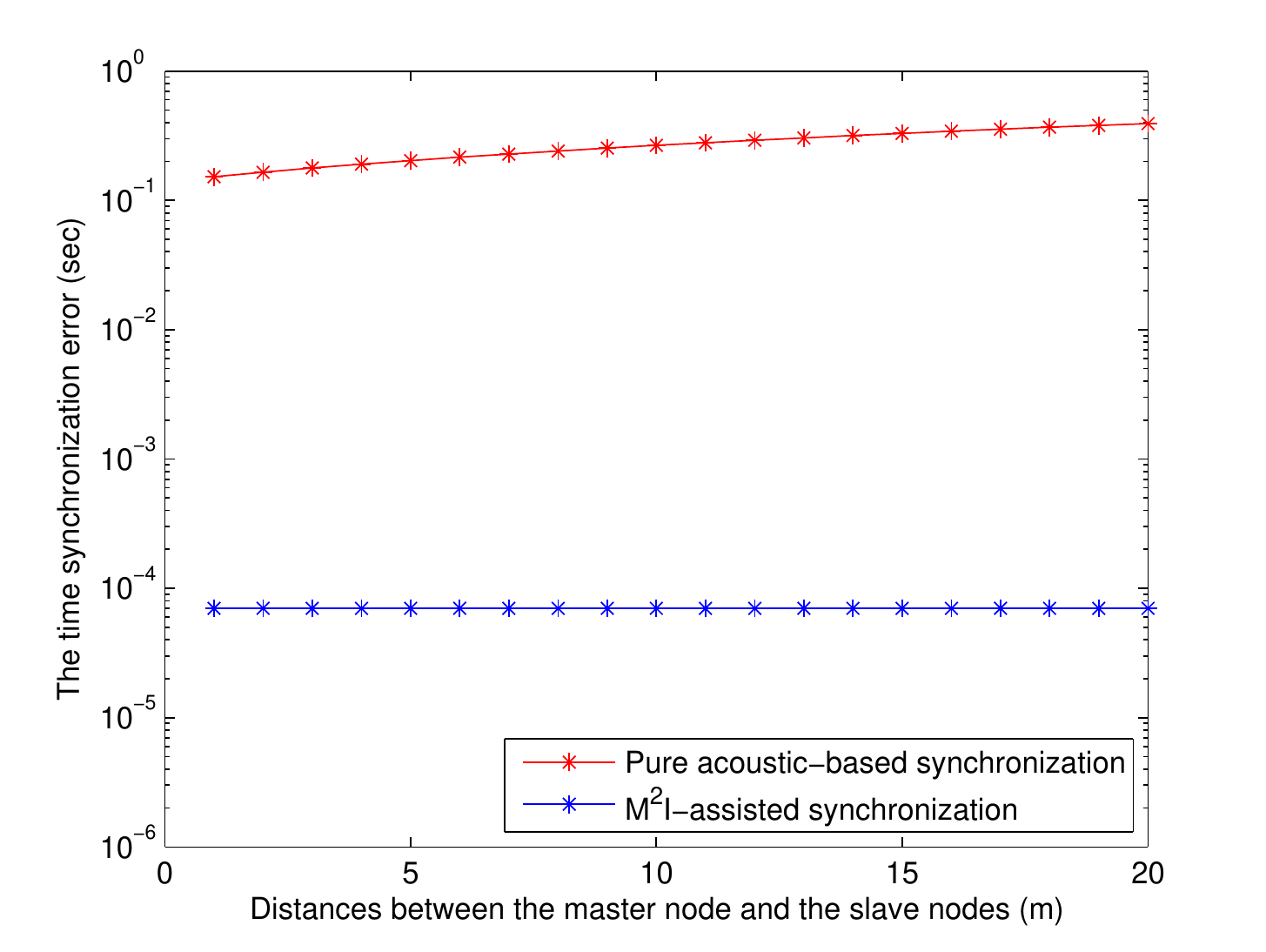}\\
  \caption{The time synchronization error of  M\(^2\)I-assisted and pure acoustic based communication.}\label{timeerror1}
\end{figure}

Then we evaluate the frequency and time synchronization error in Fig. \ref{frequencyerr} and Fig. \ref{timeerror1}. 
The operating frequency is set to be $10$ kHz for acoustic-assisted synchronization and $10$ MHz for  M\(^2\)I-assisted synchronization. 
The frequency of the crystal oscillator is $100$ kHz. 
The number of transmitting nodes including the master node is 10. 
The distance between the slave nodes and master node are set to be $20$ meters.
The packet length $L_{n,1}=100$ bits and $L_{1,n}=200$ bits. 
The bandwidth is set to be $10$ kHz for acoustic channels and $20$ kHz for  M\(^2\)I channels. 
As shown in Fig. \ref{frequencyerr}, the frequency error is evaluated by considering the variation of the frequency estimation errors.
Since the  M\(^2\)I-assisted synchronization uses higher frequency to synchronize the local hardware clock with lower frequency, the value $k$ is much larger, thus the error will be much lower than that of pure acoustic-based synchronization using low frequency acoustic waves.
As shown in Fig. \ref{timeerror1}, the time synchronization error by considering acoustic communications is extremely high since the delay of the acoustic signals enlarges the drifting time.
Shown as the blue curve, the time synchronization error can be reduced significantly once the  M\(^2\)I-assisted synchronization is used.

\section{Acoustic MIMO Module}

\vspace*{-0.1mm}
The underwater distributed MIMO is developed for either maximum SNR beamforming or space-time coding scheme. 
The long distance acoustic data transmission in the system happens after the master-slave synchronization is completed. 
In this section, the performance of the acoustic MIMO system is analyzed considering the synchronization errors found in the previous section.

The transmitted baseband signal is a train of raised cosine, bearing modulated symbols $m(t)$ \cite{chirpsignal}. 
The received signal through the acoustic channel from one transmitter can be expressed in time domain 
\begin{equation}\label{received}
  r(t)=m(t)e^{j 2\pi f} \sum^{N_{pa}}_{p=1} \frac{1}{\sqrt{P_{att}(f,d_p)}}e^{-j2\pi f \tau_p} + n(t),
\end{equation}
where $f$ is the carrier frequency and $n(t)$ is the noise. 
The time-invariant channel is considered within the synchronization time interval. 
$N_{pa}$ is the number of paths. 
$\tau_p$ is the propagation delay of the $p$-th path. 
$P_{att}(f,d_p)$ is the channel attenuation depending on the distance $d_p$ and the frequency $f$, which can be expressed as 
\begin{equation}\label{attenuation}
  P_{att}(f,d_p)=\xi_p d_p^\beta e^{\alpha(f)d_p},
\end{equation}
where $\xi$ is the scattering loss. 
$d_p^{\beta}$ is the geometric spreading loss determined by the distance $d_p$ and the spreading exponent $\beta$. 
$\alpha(f)$ is the absorption coefficient. 
By considering the distributed MIMO with $N$ perfectly synchronized transmitters, the received signal can be written as
\begin{equation}\label{perfectMIMO}
  r_N(t)=m(t)e^{j 2\pi f} \sum_{i=1}^N \sum^{N_{pa,i}}_{p=0} \frac{1}{\sqrt{P_{att,p,i}(f,d_p)}}e^{-j2\pi f \tau_{p,i}} + n(t),
\end{equation}
where the subscript $i$ indicates the $i$-th transmitter. 
By considering the uniform distributed channel delay $\tau_{p,i} \in[-\pi,\pi)$, (\ref{perfectMIMO}) can be simplified to
\begin{equation}\label{simplified}
  r'_N(t)= \sum^{N}_{i=1} A e^{j 2\pi f t} h_i e^{-j2\pi f \tau_i} + n(t),
\end{equation}
where $A$ is the amplitude of the transmitted signal. $h_i$ is the channel envelope that follows the PDF below \cite{model} :
\begin{align}\label{PDF}
p_{h_i}(z)=4 \pi^2 z \int_0^{\infty} \prod_{p=1}^{N_{pa,i}} J_0(2 \pi |h_{p,i}|x) J_0(2 \pi z x) J_0(2 \pi h_{0,i}x)x dx,
\end{align}
where $h_{p,i}$ is the channel attenuation of the $p$-th path for the transmitter $i$ that has
\begin{equation}\label{jattenuation}
  h_{p,i}= \frac{1}{\sqrt{P_{att,p,i}(f,d_p)}}.
\end{equation}
$J_0(x)$ is the Bessel function of the zeroth order and the first kind where
\begin{equation}\label{J0}
  J_0(x)=\frac{1}{2 \pi} \int^{2 \pi}_0 e^{j x cos\beta} d \beta.
\end{equation}
By considering the perfect CSI information, the channel envelope $h_i$ and the delay $\tau_i$ are known.

As mentioned before, the underwater distributed MIMO is developed for either maximum SNR beamforming or space-time coding scheme. 
First, we analyze the performance of the beamforming by considering  M\(^2\)I/acoustic-based synchronization. 
The received signal after the phase control process can be written as
\begin{equation}\label{phasecontrol}
  r_{N,BF}(t)= \sum^{N}_{i=1} A e^{j 2\pi f t} h_i e^{-j 2\pi f \tau_i} e^{j \phi_i} + n(t),
\end{equation}
where $\phi_i$ is the phase control term at the $i$-th transmitter. The objective of the beamforming is to align the phase and maximize the SNR at the receiver side. 
The SNR can be expressed as SNR= $\frac{A^2 \left|\mathbf{h}^\top \mathbf{v}_{\phi} \right|^2}{\sigma^2}$.
Therefore, the optimal phase control vector $\mathbf{v}_{\phi}$ can be obtained by maximizing the SNR, which is
\begin{align}\label{optimization}
\begin{split}
\max_{\mathbf{v}_{\phi}}\quad SNR,
\end{split}
\end{align}
where
\begin{align}\label{vectors}
\begin{split}
\mathbf{h} \triangleq [h_1e^{j 2\pi f (t-\tau_1)},\ h_2e^{j 2\pi f (t-\tau_2)},\ ...\ h_Ne^{j 2\pi f (t-\tau_N)}], \\
\mathbf{v}_\phi \triangleq [e^{j \phi_1},\ e^{j \phi_2},...\ e^{j \phi_N}], \quad \quad \sigma^2 \triangleq \mathbb{E}\{|n(t)|^2\}.\\
\end{split}
\end{align}
By considering the optimal phase control, the channel delay can be compensated by the phase control and the SNR at the received side can be maximized as
\begin{equation}\label{BFSNR}
  SNR_{BF}=\frac{A^2 \left|\sum_{i=1}^N h_i\right|^2}{\sigma^2}.
\end{equation}

According to the synchronization error analysis presented in Section IV, the received signal can be expressed by considering the frequency and time errors
\begin{equation}\label{phasecontrol1}
  r_{N,BF,\epsilon}(t)= \sum^{N}_{i=1} A h_i e^{j 2\pi (f+\epsilon_{i,f}) (t-\tau_i+\epsilon_{i,t})} e^{j \phi_i} + n(t),
\end{equation}
where the $\epsilon_{i,t}$ and $\epsilon_{i,f}$ are respectively the time and frequency error of the $i$-th node. 
$i$=$1$ indicates the master node. 
Therefore we have $\epsilon_{1,t}=0$ and $\epsilon_{1,f}=0$. 
The phase controlled SNR by considering the time and frequency error can be written as
\begin{equation}\label{errormax}
  SNR_{BF,\epsilon}(t)=\frac{A^2 \left|\sum_{i=1}^N h_i e^{j 2\pi (f+\epsilon_{i,f}) (t+\epsilon_{i,t})}\right|^2}{\sigma^2}.
\end{equation}

Similarly, the maximum SNR and error considered SNR for the space-time coding can be represented by
\begin{align}\label{STBCSNR}
\begin{split}
SNR_{STBC}&=\frac{A^2 \sum_{i=1}^N h_i^2}{\sigma^2}, \\
SNR_{STBC,\epsilon}(t)=&\frac{A^2 \sum_{i=1}^N \left|h_i^2 e^{j 2\pi (f+\epsilon_{i,f}) (t+\epsilon_{i,t})}\right|}{\sigma^2}.
\end{split}
\end{align}

Due to the relative clock drift, the SNR presented in (\ref{BFSNR}) and (\ref{STBCSNR}) will decrease with time. 
To maintain the communication, the SNR is required to be greater than a threshold $\eta$. 
Once the SNR is about to be lower than the threshold, another round of synchronization is required. 
The effective communication time is defined as the duration between two adjacent rounds of synchronization that can be used to transmit useful information. Therefore, the effective communication time for the beamforming and space-time coding $t_{BF,SNR}$, $t_{STBC,SNR}$ according to the SNR is constrained by
\begin{align}\label{communicationtimeBF}
\begin{split}
&t_{BF,SNR}=\min_{t} \ t \\
s.t. \quad &SNR_{BF,\epsilon}(t) < \eta, \\
\end{split}
\end{align}
\begin{align}\label{communicationtimeSTBC}
\begin{split}
&t_{STBC,SNR}=\min_{t} \ t \\
s.t. \quad &SNR_{STBC,\epsilon}(t) < \eta. \\
\end{split}
\end{align}
Since the channel model is assumed to be quasi-static, the effective communication time cannot be greater than the coherence time (the time that channel remains static) defined in \cite{coherence}, which is
\begin{equation}\label{coherencetime}
  T_c=\sqrt{\frac{9}{16 \pi f_d^2}} \approx \frac{0.423}{f_d}=\frac{0.423}{a f},
\end{equation}
where $f_d$ is the Doppler shift and $a$ is the Doppler scaling factor.
For the underwater distributed MIMO system, the transmitters have to redo the synchronization and CSI estimation if either the $SNR<\eta$ or the effective communication time oversteps the coherence time. 
Therefore, the effective communication time by considering both the SNR and the coherence time can be obtained by
\begin{equation}\label{BFctime}
  t_{BF}=\text{argmin}\{t_{BF,SNR},T_c\},
\end{equation}
and
\begin{equation}\label{STBCctime}
  t_{STBC}=\text{argmin} \{t_{STBC,SNR},T_c\}.
\end{equation}
\vspace{-0.5mm}The efficiency of the underwater distributed MIMO system can be evaluated by calculating the throughput of data. 
By considering the CSI estimation time and synchronization time, the upper bound of the throughput by using beamforming and space-time coding can be respectively written as
\begin{align}\label{throughput}
\begin{split}
  Tp^{BF}=\frac{t_{BF}C_{BF}}{\sum_{i=2}^N \Delta t_i +t_{CSI}+t_{BF}},\\
  Tp^{STBC}=\frac{t_{STBC}C_{STBC}}{\sum_{i=2}^N \Delta t_i +t_{CSI}+t_{STBC}},
\end{split}
\end{align}
where $t_{CSI}$ is the time of the broadcast from the base station for the CSI estimation calculated by:
\begin{equation}\label{CSI}
  t_{CSI} = \frac{L_{CSI}}{B_{ac}} + \frac{d_{max}}{v_{ac}}.
\end{equation}
$L_{CSI}$ is the total packet length delivered for the CSI estimation. 
$B_{ac}$ is the bandwidth of the acoustic channel and $v_{ac}$ is the propagation speed of the acoustic signals. 
$d_{max}$ is the maximum distance between the based station and the sensor nodes. 
$C_{BF}$ and $C_{STBC}$ in (\ref{throughput}) are respectively the channel capacity of the distributed MIMO by using beamforming and space-time coding techniques. 
By considering the frequency synchronization error, the SNR is not a constant so that the channel capacity varies with the time as well. To calculate the upper bound presented in (\ref{throughput}), the maximum SNR obtained at the beginning of the communication can be used and then $C_{BF}$ and $C_{STBC}$ can be expressed as
\begin{align}\label{channelcapacity}
\begin{split}
  C_{BF}&=\text{log}_2 \left[ \text{det}\left( \mathbf{I}_N + \frac{SNR_{BF,\epsilon}(t=0)}{N_{b} }\mathbf{H} \mathbf{H}^*\right)\right],\\
  C_{STBC}&=\text{log}_2 \left[ \text{det}\left( \mathbf{I}_N + \frac{SNR_{STBC,\epsilon}(t=0)}{N_{b} }\mathbf{H} \mathbf{H}^*\right)\right],
\end{split}
\end{align}
where $N_b$ is the number of the base stations. 
$\mathbf{I}_N$ denotes the identity matrix of size $N$. 
$\mathbf{H}$ is the $N \times N_b$ channel matrix.

\section{Numerical Evaluations}

\begin{figure}
  \centering
  \includegraphics[width=2.2in]{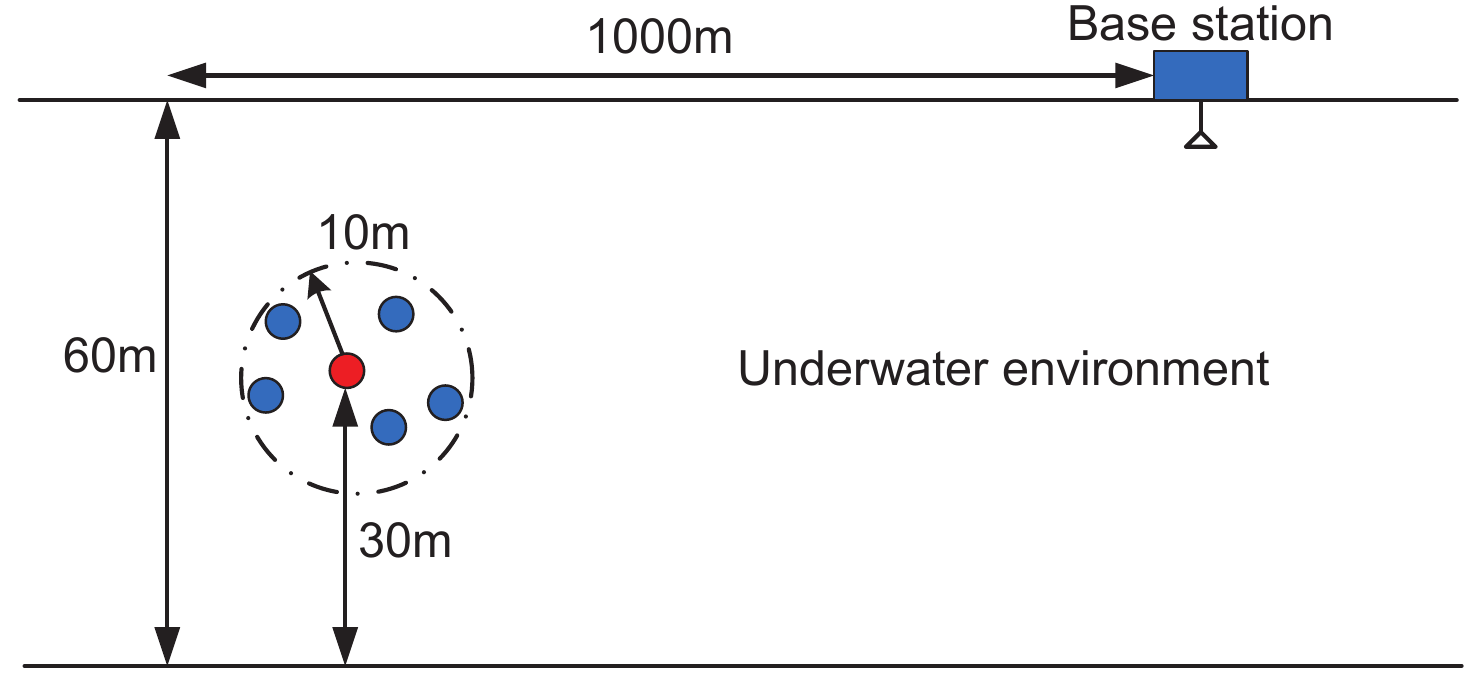}\\
  \caption{The geometry of the underwater distributed MIMO system.}\label{parameters}
\end{figure}

In this section, we evaluate the performance of the proposed underwater distributed MIMO systems by comparing the pure acoustic MIMO system and hybrid  M\(^2\)I-acoustic MIMO systems. 
The geometry of the underwater distributed MIMO system is considered as Fig. \ref{parameters}. 
The transmitting power from each transmitter is set to be $10$ mW. 
The power of the noise is $9.81\times10^{-3}$ mW \cite{noise}. 
The  M\(^2\)I transmitters operate at $10$ MHz frequency with $20$ kHz bandwidth. 
The acoustic transmitters operate at $10$ kHz frequency with $10$ kHz bandwidth.

\begin{figure}
  \centering
  \includegraphics[width=2.6in]{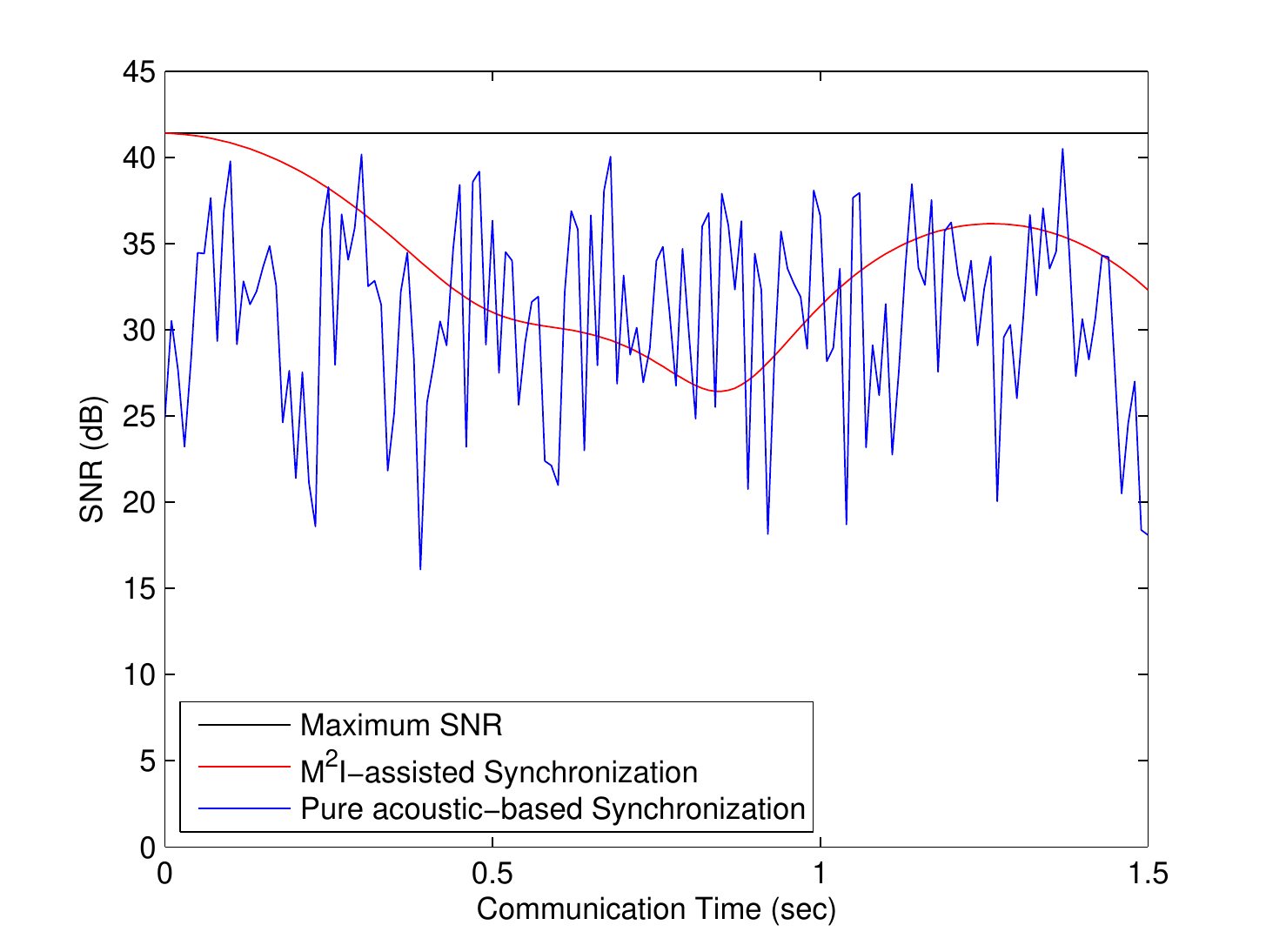}\\
  \caption{The SNR vs. time of beamforming.}\label{SNRBF}
\end{figure}

\begin{figure}
  \centering
  \includegraphics[width=2.6in]{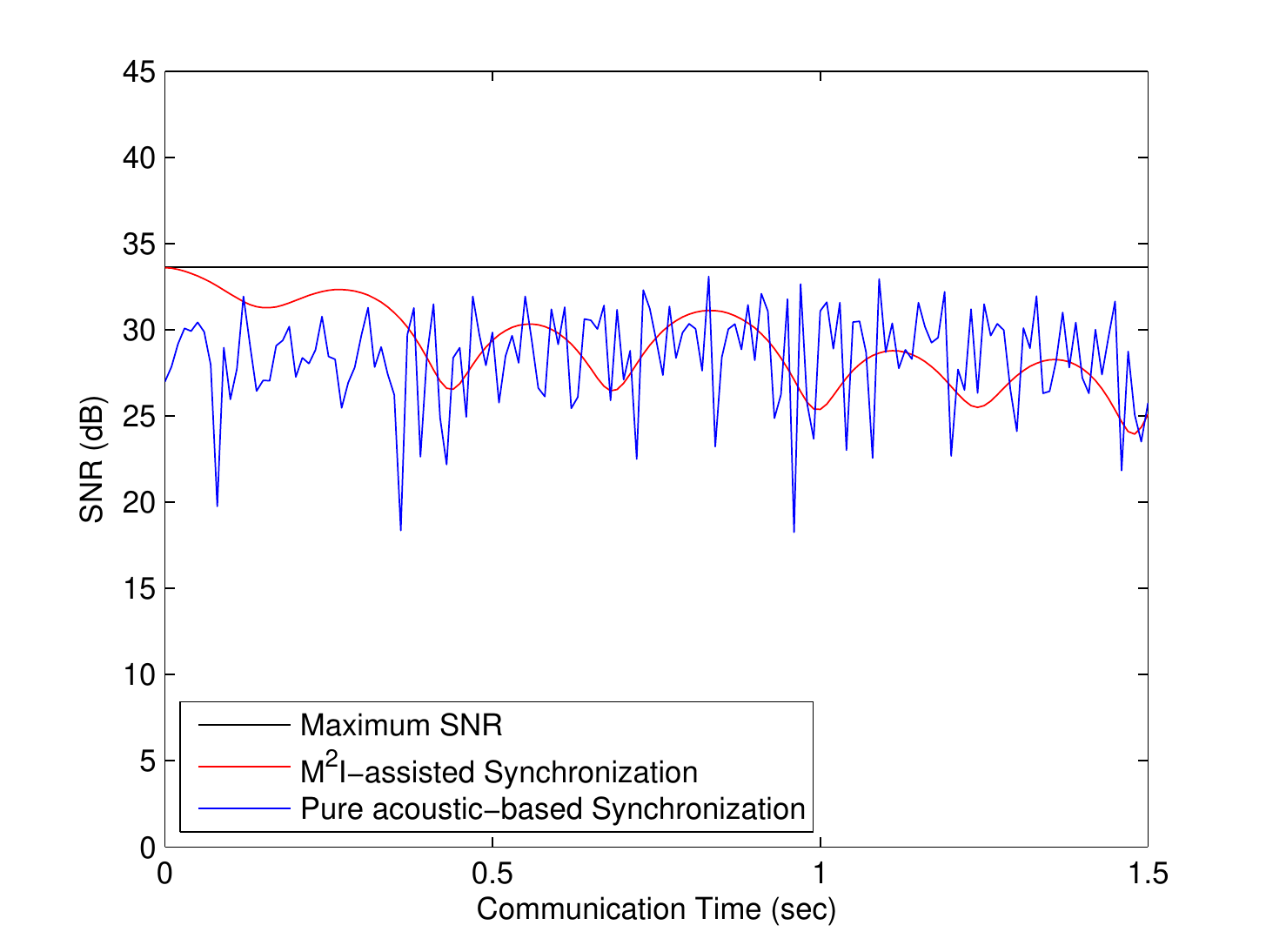}\\
  \caption{The SNR vs. time of space-time coding.}\label{SNRSTBC}
\end{figure}
The SNRs vs. time of using beamforming and space-time coding scheme are evaluated in Fig. \ref{SNRBF} and Fig. \ref{SNRSTBC}, respectively. 
The SNR is calculated for the first 1.5 seconds after completing the synchronization. 
Five slave nodes are randomly deployed around a master node within $10$ meters range. 
The black horizontal line is the maximum SNR by considering the perfect time and frequency synchronization.
The result of using  M\(^2\)I-assisted synchronization is shown as the red curve. 
At the beginning $t=0$, the SNR is very close to the upper bound since the  M\(^2\)I transceivers can provide very accurate synchronization due to small delay and large bandwidth.  
However, the SNR decreases as the time increases and becomes random since the phases are not aligned after a certain time.
Shown as the blue curve, the SNR of using pure acoustic-based synchronization randomly varies since the acoustic communication cannot provide accurate synchronization. 
The phases cannot be aligned even at the beginning of the communication.

\begin{figure}
  \centering
  \includegraphics[width=2.6in]{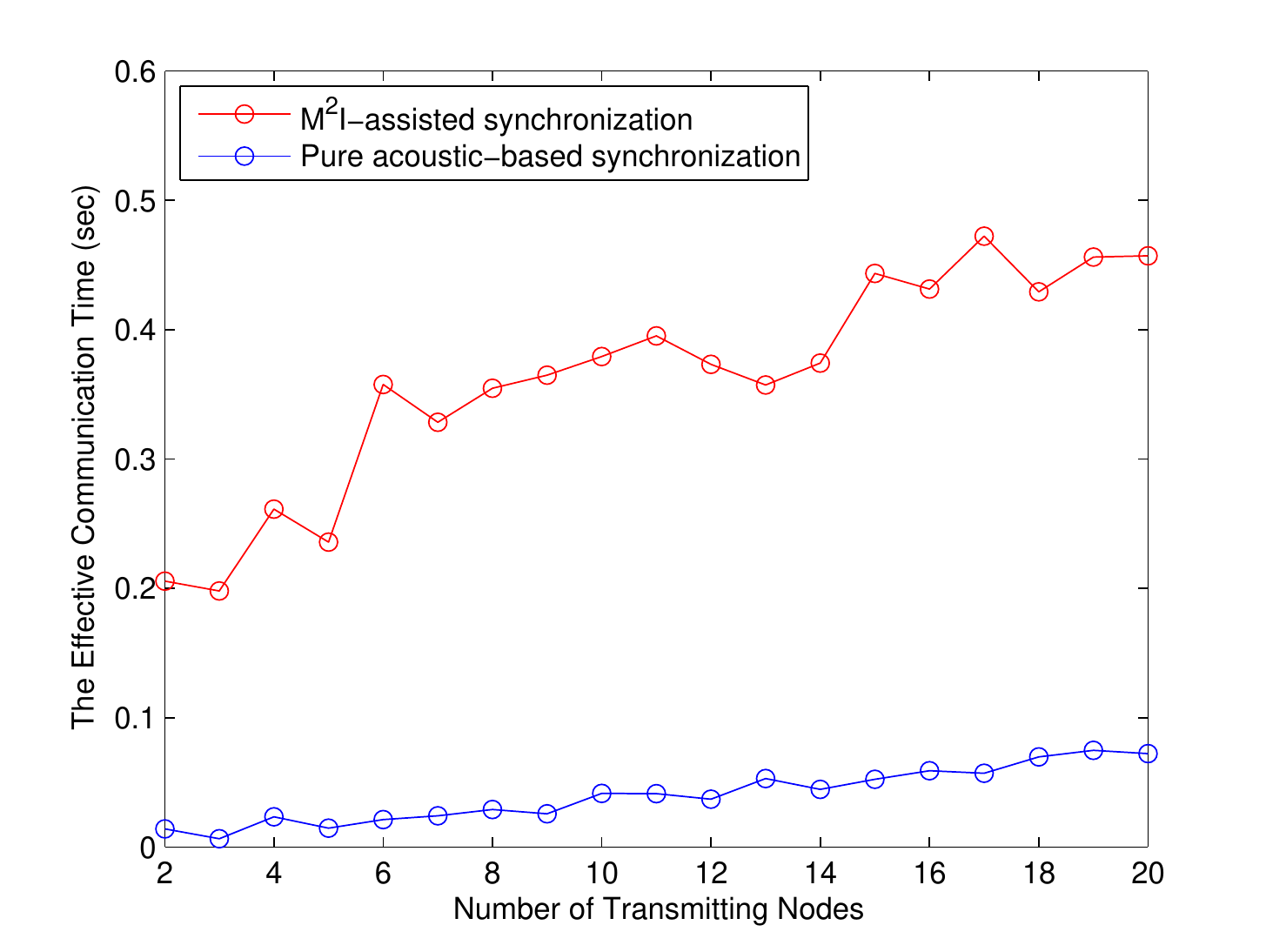}\\
  \caption{The effective communication time of beamforming.}\label{tcBF}
\end{figure}

\begin{figure}
  \centering
  \includegraphics[width=2.6in]{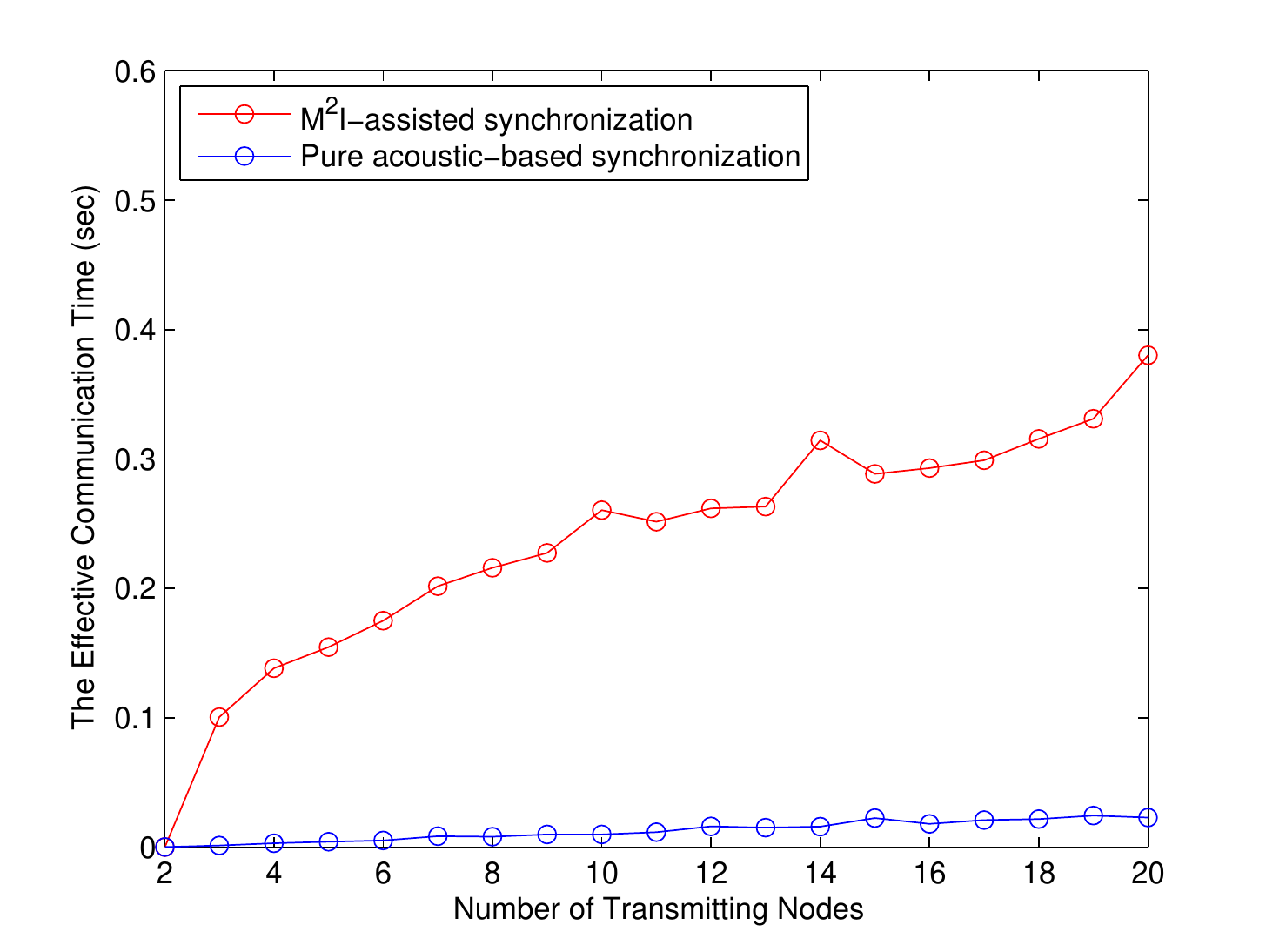}\\
  \caption{The effective communication time of space-time coding.}\label{tcSTBC}
\end{figure}

The effective communication time in (\ref{BFctime}) and (\ref{STBCctime}) is evaluated in Fig. \ref{tcBF} and Fig. \ref{tcSTBC}. 
The threshold of minimum SNR is set as $\eta=25$ dB.
When SNR is below 25 dB, another round of synchronization is required to maintain the communication. 
In this evaluation, we let the number of transmitting nodes increase from 2 to 20. 
For each number of transmitting nodes, we randomly deploy the slave nodes for $100$ times and calculate the average effective communication time. 
The effective communication time increases with the number of transmitting nodes increases since the total transmitting power is larger if more nodes are used. However, the effective communication time does not increase monotonically since it takes more time to synchronize more nodes. 
As a result, the synchronization error becomes larger. 
Moreover, the randomness of the nodes' deployment also influences the effective communication time. 
The result shows that the distributed MIMO system can have more effective communication time if the  M\(^2\)I-assisted synchronization is used.

\begin{figure}
  \centering
  \includegraphics[width=2.6in]{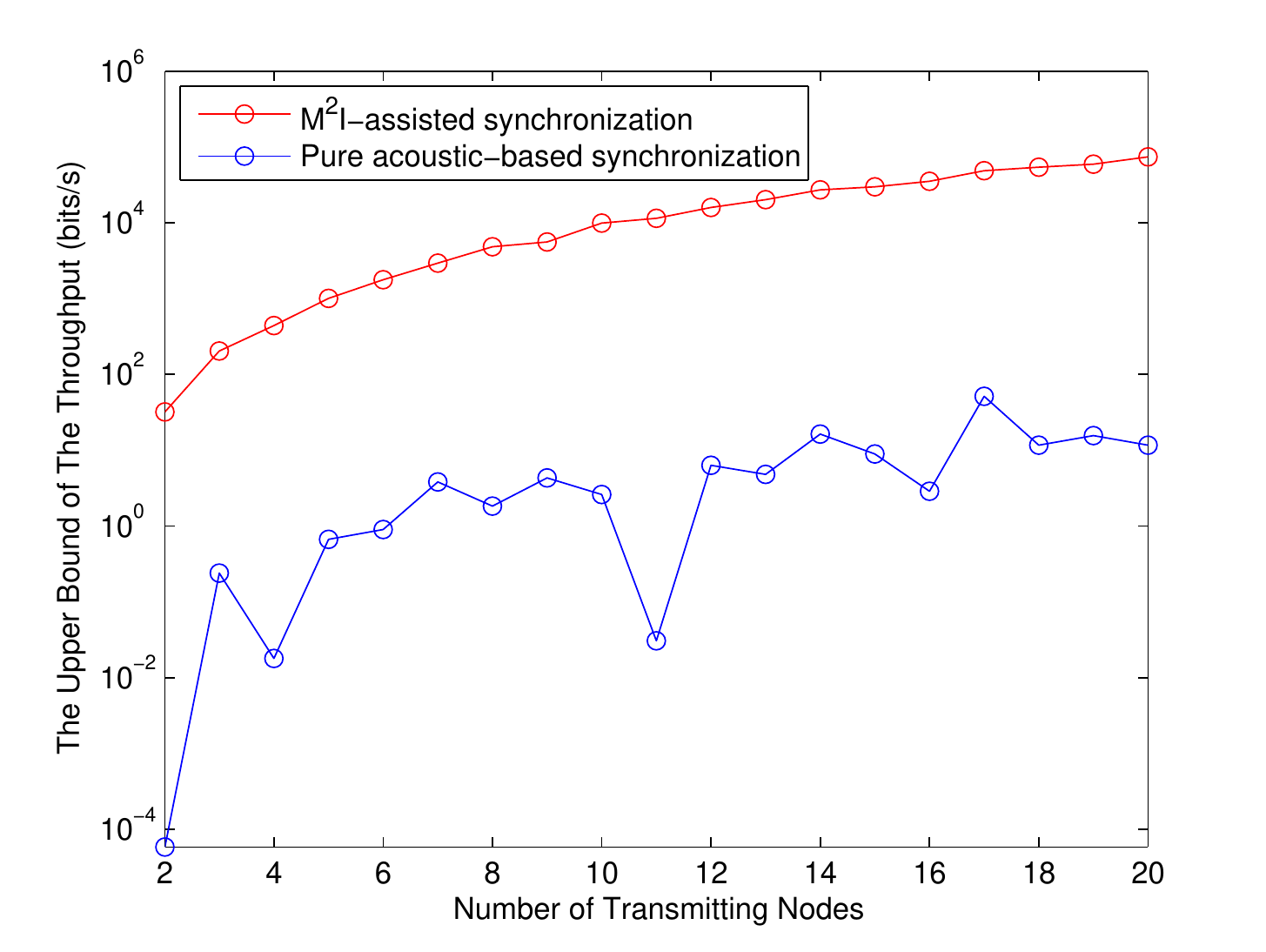}\\
  \caption{The upper bound of the throughput using beamforming techniques.}\label{TpBF}
\end{figure}

\begin{figure}
  \centering
  \includegraphics[width=2.6in]{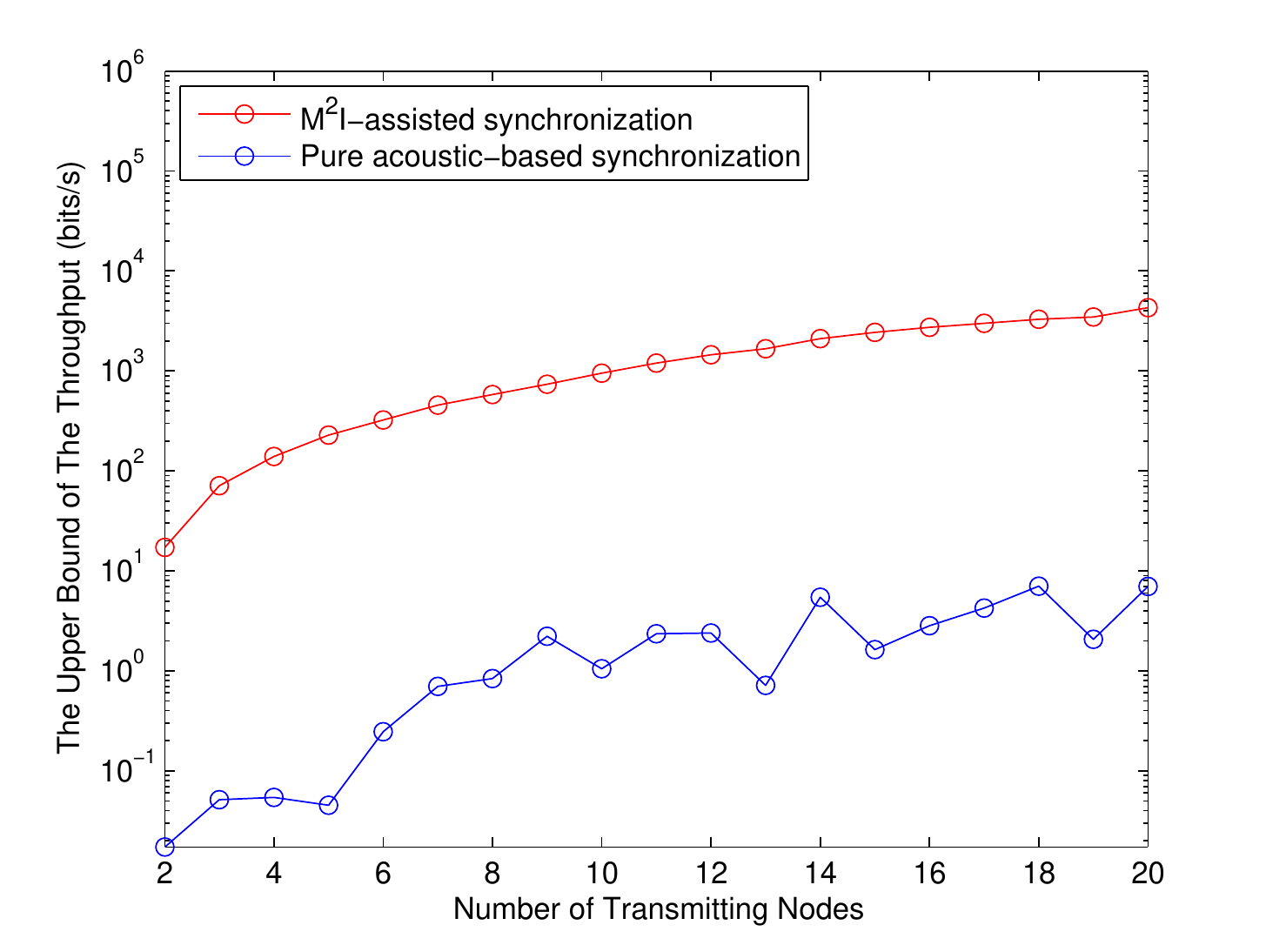}\\
  \caption{The upper bound of the throughput using space-time coding techniques.}\label{TpSTBC}
\end{figure}

Finally, we calculate an upper bound of the throughput in Fig. \ref{TpBF} and Fig. \ref{TpSTBC}. 
Compared to the distributed MIMO system synchronized by acoustic communications, the  M\(^2\)I synchronized systems have shorter synchronization time and longer effective communication time. 
Therefore, the throughput of distributed acoustic MIMO system can be enhanced by using  M\(^2\)I-assisted synchronization.
The numerical results show that the throughput upper bound of hybrid acoustic and M\(^2\)I distributed MIMO system exceeds that of pure acoustic distributed MIMO system two orders of magnitude in both beamforming and STBC scenarios.

\begin{figure}
  \centering
  \includegraphics[width=\linewidth]{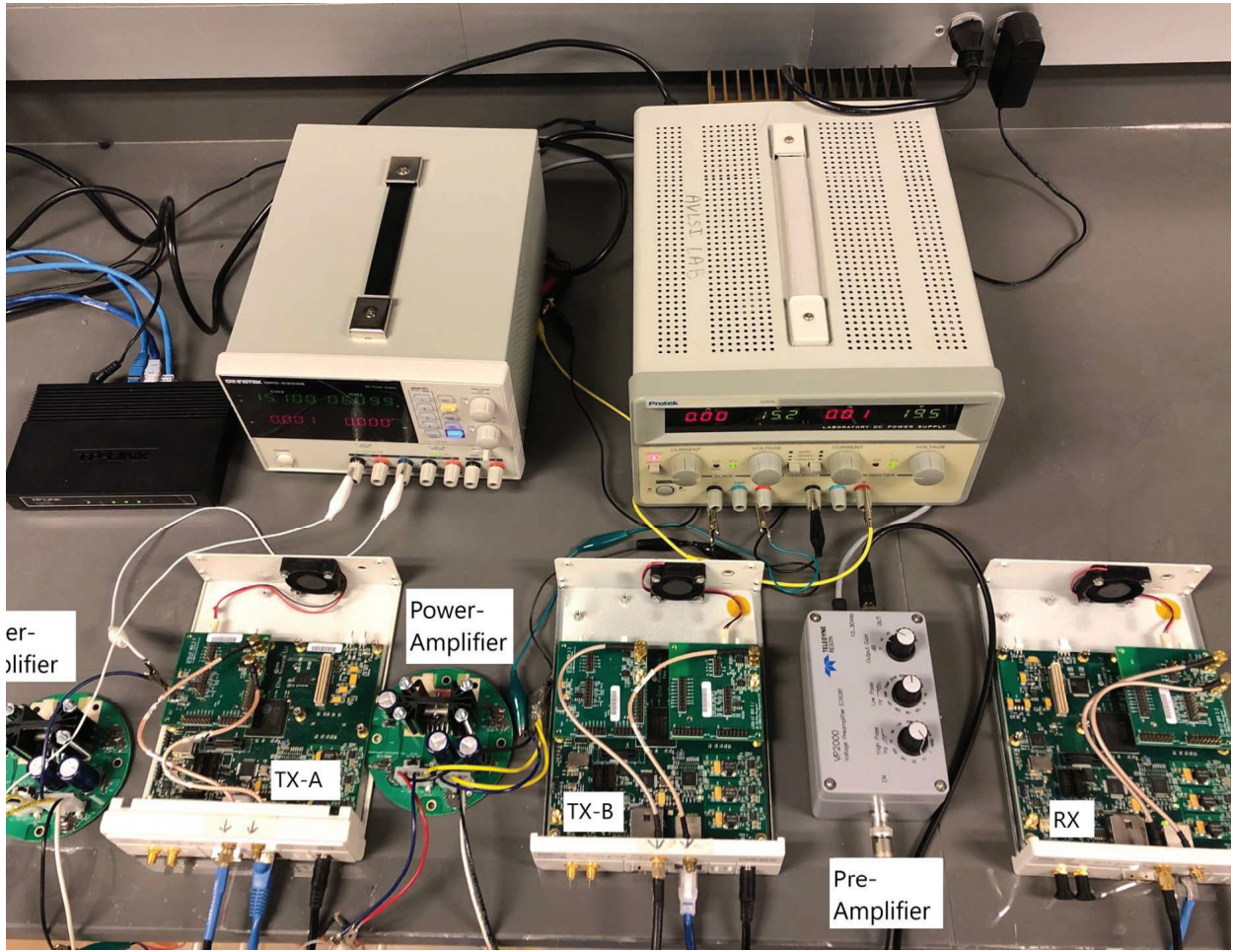}\\
  \caption{SDAMMS testbed setup based on USRP N210.}\label{exp}
\end{figure}

\begin{figure*}[!htb]
\minipage{0.4\textwidth}
  \includegraphics[width=\linewidth]{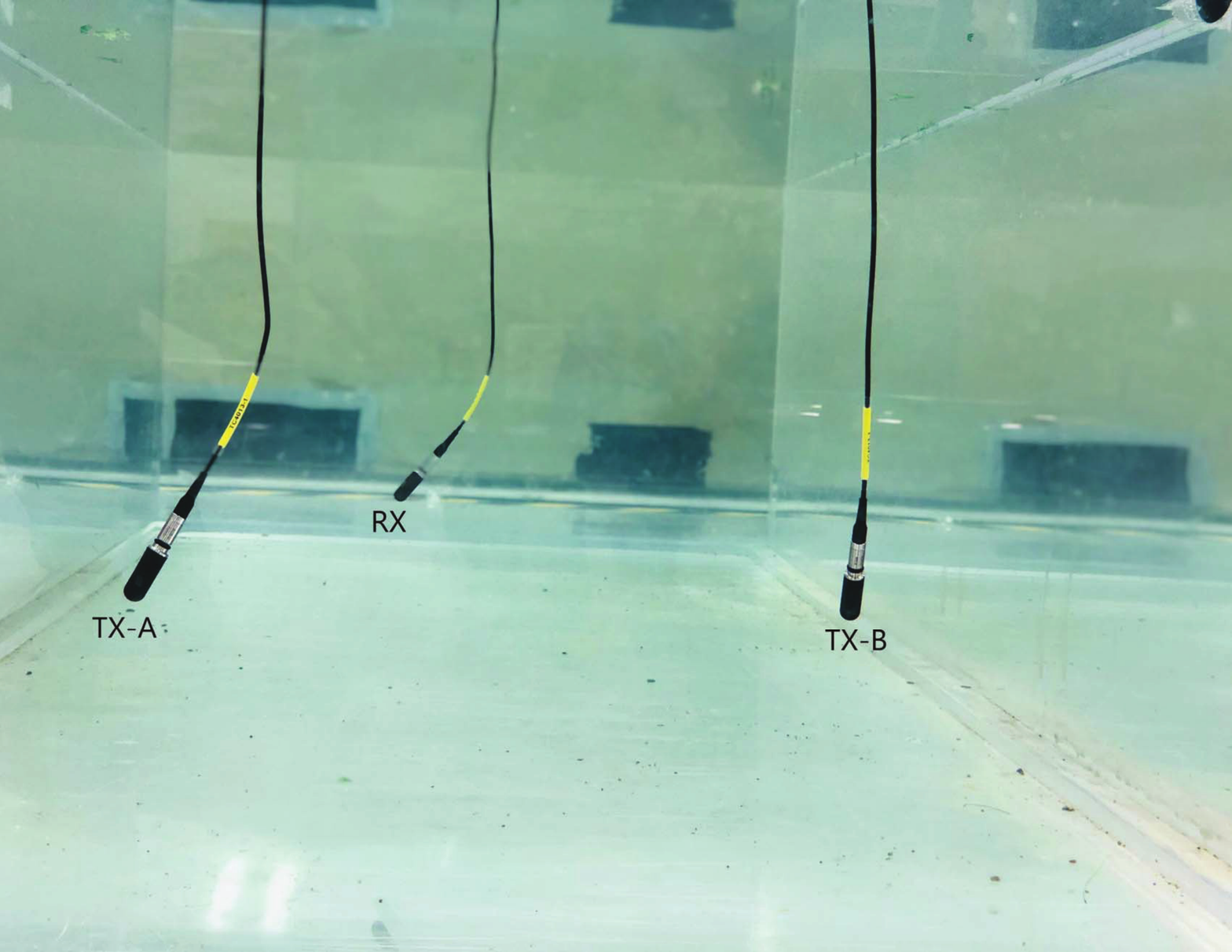}
  \caption{Acoustic Hydrophones deployed in water tank.
  }\label{Acoustic}
\endminipage\hfill
\minipage{0.455\textwidth}
  \includegraphics[width=\linewidth]{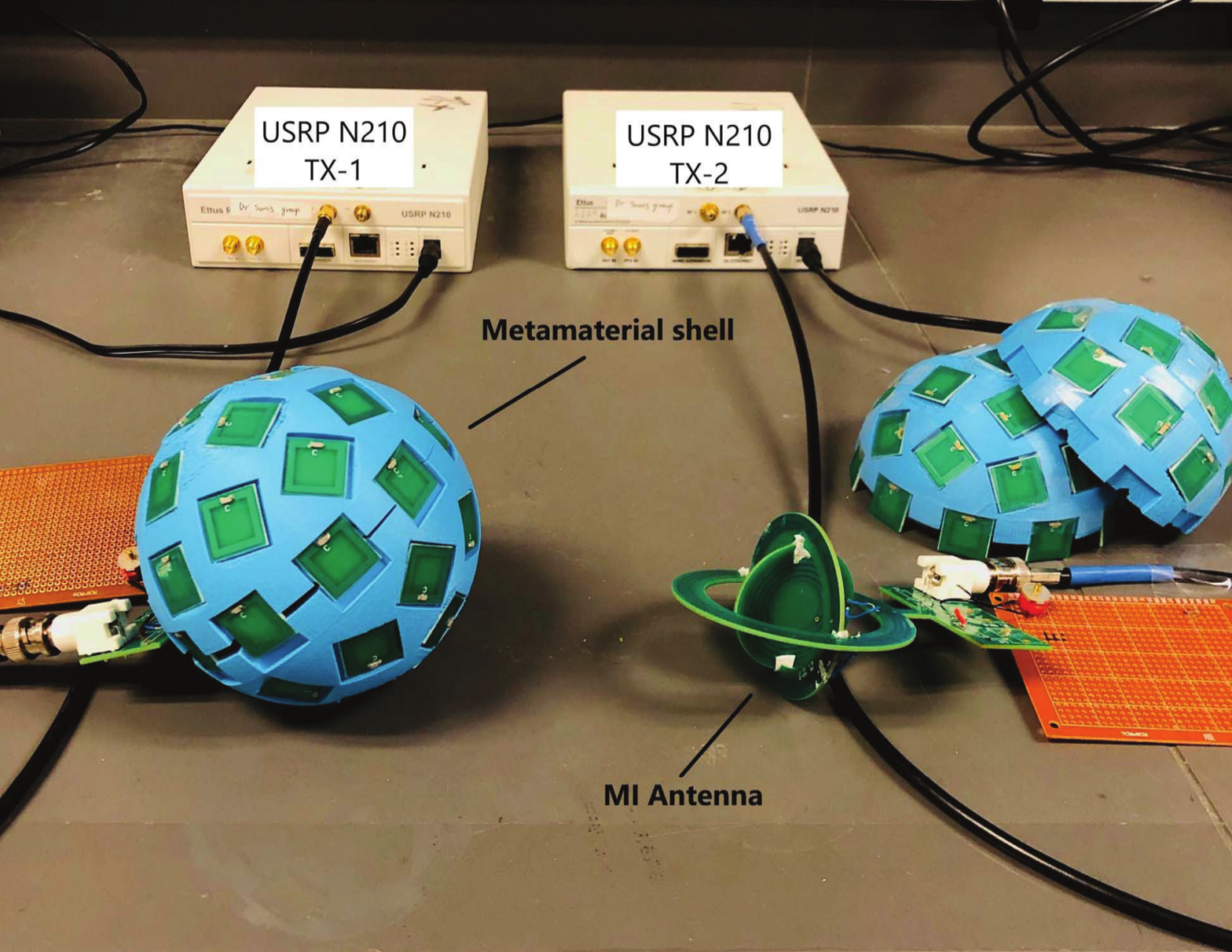}
  \caption{M\(^2\)I antennas used for Synchronization.}\label{MI}
\endminipage\hfill
\end{figure*}

\begin{figure*}
\centering
\includegraphics[width=6.5in]{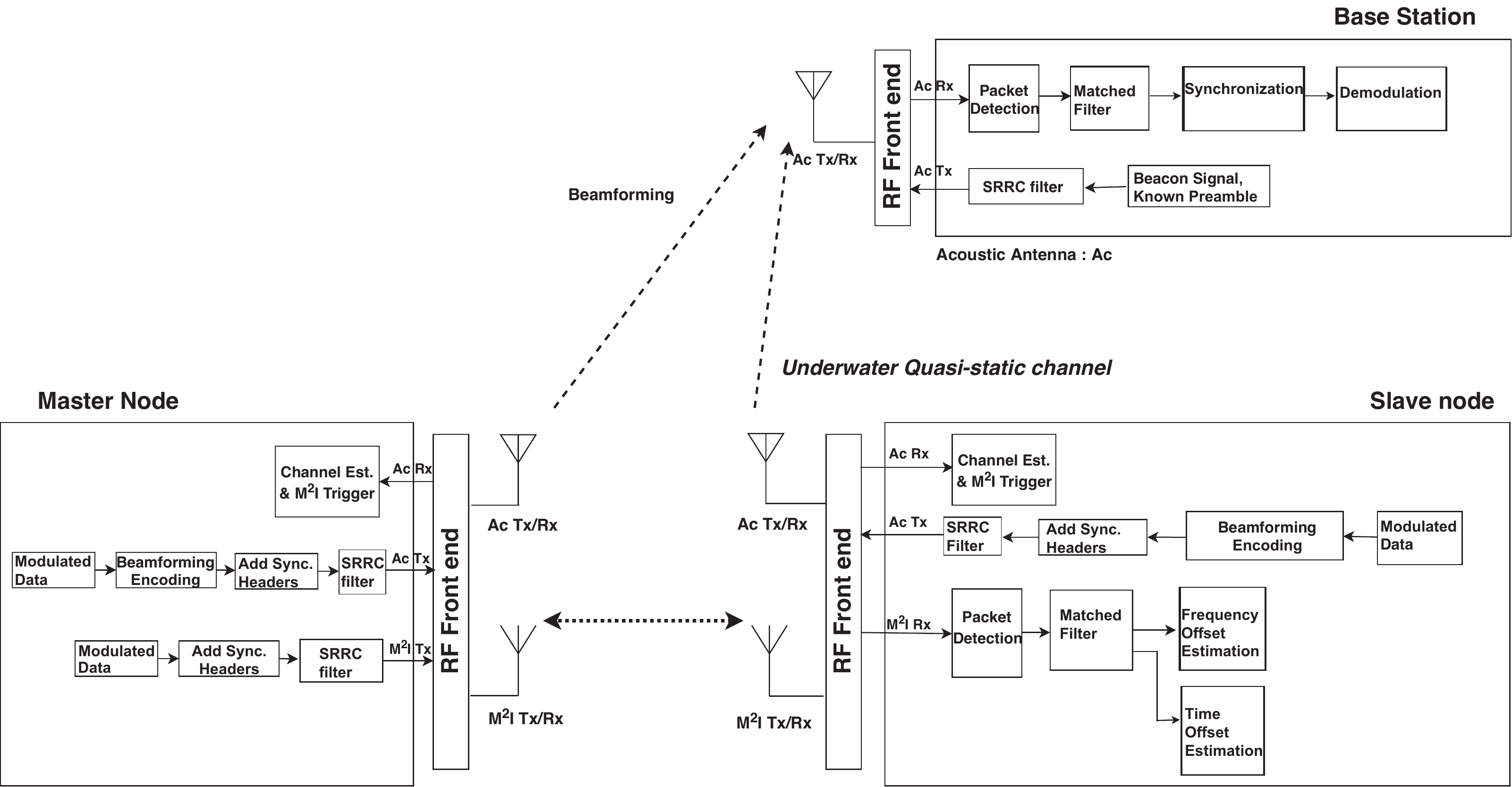}\\
 \caption{Block diagram of underwater distributed  hybrid M\(^2\)I and acoustic MIMO. }\label{block}
\end{figure*}

\section{Testbed Implementation and Experimental Evaluations}
The capabilities of the proposed system architecture are demonstrated using software-defined testbed in a lab-controlled tank environment. 
The setup is inspired by the platform established in \cite{SDR, disbeam,Dmimo}. 
Our proposed architecture is implemented in software-defined acoustic and  M\(^2\)I modems (SDAMM) that are built in-house. 
In this section, we analyze the performance of SDAMMs as well as the system implementing configuration. 
We use 3 SDAMMs in our experiments, two transmitters and one receiver to established an $2\times 1$ hybrid acoustic and M\(^2\)I MIMO System.

\subsection{Software-Defined Testbed Setup}
All three SDAMMs are based on the Universal Software Radio Peripheral (USRP) N210 platform and are equipped with LFTX and LFRX daughter-boards that can operate from 0 to 30 MHz, which is within the operating frequency band of M\(^2\)I antennas.
Experiment testbed setup is shown in Fig.\ref{exp}. 

The two transmitting SDAMMs are connected to a host-PC via a Gigabit Ethernet switch. 
Each SDAMM is then connected to one acoustic hydrophone and one  M\(^2\)I antenna in the following way: The LFTX daughter-board is connected to a linear wideband power amplifier, Benthowave BII-5002. 
Through the power amplifier, the SDAMM is connected to Teledyne RESON TC4013 acoustic transducer deployed in the water tank, which is shown in Fig. \ref{Acoustic}.
As seen in Fig. \ref{MI}, the LFTX daughter-board on each transmitting SDAMM is also connected to a M\(^2\)I antenna, which is consist of an tri-directional MI antenna and a metamaterial shell.  

The receiving side of the testbed can also be seen in Fig. \ref{exp}, the received signal from the Teledyne RESON TC4013 acoustic transducer is first passed through the voltage preamplifier and adjustable bandpass filter, Teledyne VP2000.
Then the signal gets down-converted to the baseband for further processing by the LFRX daughterboard. 
This USRP device is also connected to a host PC via Ethernet. 
GNU Radio companion and MATLAB R2018  are run on the host PCs for signal generation and processing. 
GNU Radio companion is an open source tool to create and analyze signal flow graphs used for baseband signal processing. 
On the transmitter side, the required signals and packets are generated using GNU Radio and MATLAB then transferred to the SDAMMs using Universal Hardware Driver (UHD). 
Similarly, at the receiver, the baseband signal from the LFTX daughterboard is given to GNU Radio for further signal processing which is done using custom digital signal processing blocks.

All the underwater MIMO experiments are conducted in a in-lab water tank of size 8 ft x 2.5 ft x 2 ft.
The transmission gains and power for the transmitters is kept constant throughout the experiments, the carrier frequency ($f_c$) of  M\(^2\)I antennas is set as 30MHz and of acoustic hydrophones is set as 100kHz.

\subsection{Hybrid MIMO System Implementation}
In this section, we discuss the proposed system implementation detail using GNU radio companion and MATLAB R2018. A complete overview of the system implementation can be seen in Fig.\ref{block}.

\subsubsection{ M\(^2\)I-assisted Synchronization Module}

The overview of synchronization process can be seen in Fig. \ref{block}.
The master node sends data packets with synchronization preamble to the slave node using  M\(^2\)I antenna. 
Data is modulated using Binary Phase-shift Keying (BPSK) modulation scheme and the synchronization header consists of two repeated 127 bits short PN sequences and two repeated 1024 bit long PN sequences, these short and long sequences will be used for coarse and fine frequency offset estimation at the slave node. 
A chirp signal is also added as a header to the data packets. 
Chirp signals have excellent robustness against multipath, Doppler spread and also have useful correlation characteristics.
Hence, a chirp signal is used for start of packet detection and time offset estimation at the slave node. 
The modulated data and synchronization headers also pass through a square-root-raised-cosine (SRRC) filter before transmission.

For frequency synchronization and Carrier Frequency Offset (CFO) estimation using the repeating preambles, a maximum likelihood method is utilized according to 802.11a short preamble method in \cite{cfo, maxlike}, where multiple small preambles are transmitted in order to estimate the CFO.
On reception, cross correlation is performed and received packet detection is done using Chirp signal.
For time synchronization, the initial time delay/time offset can be detected using cross correlation to find a rough alignment (within a sample) of two signals, which is accomplished with a custom ``sample offset" block in GNU radio.

\subsubsection{Acoustic MIMO Module}

\begin{figure}
  \centering
  \includegraphics[width=3in]{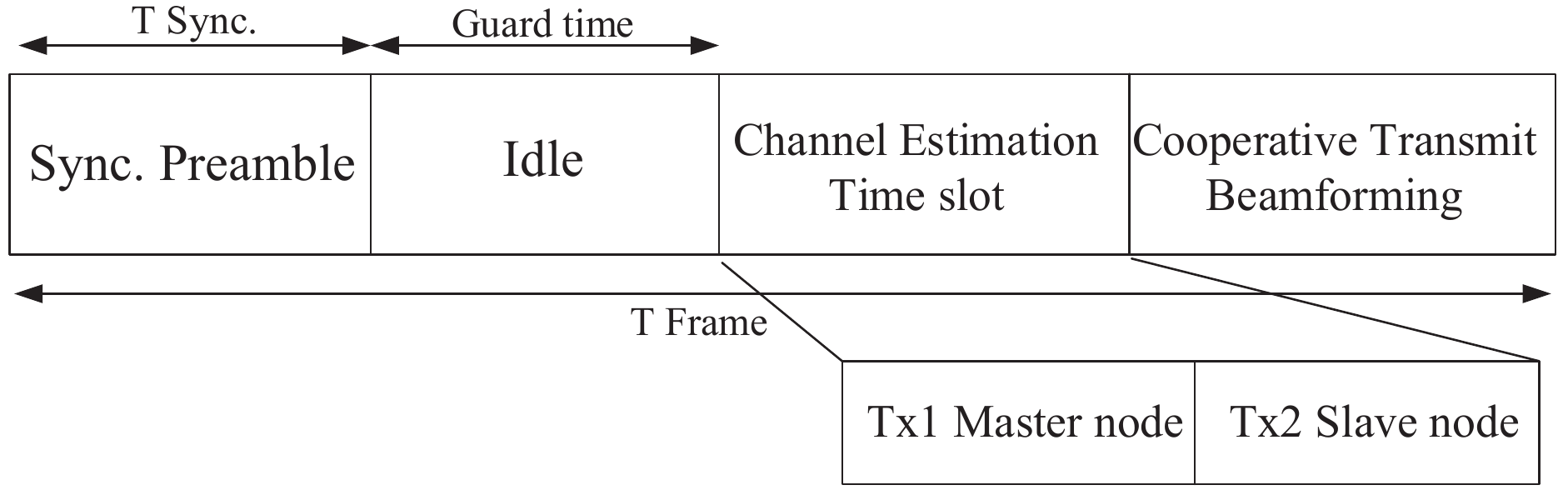}\\
  \caption{The packet structure for transmitter beamforming.}\label{bfpacket}
\end{figure}

\subsubsection*{Beamforming Transmission}
Block diagram for acoustic MIMO data transmission can also be seen in Fig.\ref{block}. 
According to our discussion in section III-B, for the maximum SNR beamforming scheme, the BS transmits a beacon signal and a known preamble which is utilized at the transmitters for channel estimation.
The preamble used here is a 127 bit PN sequence. 
For beamforming data towards the BS, the transmitters utilize BPSK modulation and data is multiplied by the phase inverse of their respective channels to the BS. 
The pack structure used for transmitter beamforming is shown in Fig. \ref{bfpacket}.
Synchronization is still needed between the transmitters and BS and hence synchronization headers are added. 
The modulated data and synchronization headers pass through a square-root-raised-cosine (SRRC) filter and are sent to the Acoustic hydrophones. 
The BS which is the receiver hydrophone implements a matched filter and data demodulation is performed post-synchronization.

The parameters configurations are described as follows: Square-root-raised-cosine filters used in the experiments have pulses of duration $T_S$ = 1.3107 ms, roll-off factor $\alpha$= 0.3. 
Modulated symbols are also power normalized and interpolated with 4 samples per symbol for upsampling. 
The sampling rate on GNU radio companion is set to 5 MHz for  M\(^2\)I transmissions and 195.312 kHz for acoustic transmissions. 
The chirp signal which is used as preamble is generated with the following parameters: Initial frequency of 0, Target frequency of 10, Target Time of 10, Sweep Time of 100, Sample Rate of 50, Samples Per Frame 500 and consists of 500 bits.

\begin{figure}
  \centering
  \includegraphics[width=3in]{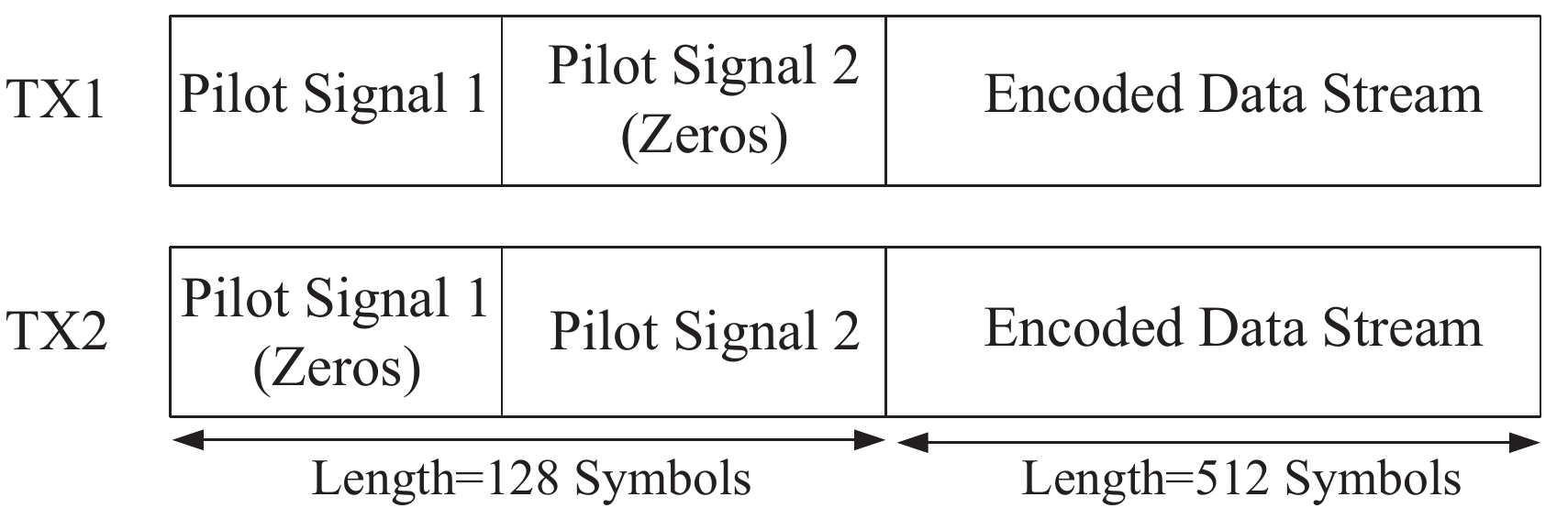}\\
  \caption{The frame that is being transmitted from TX1 and TX2 in Alamouti coding scheme.}\label{Alapacket}
\end{figure}

\subsubsection*{Space Time Coding Transmission}
Alamouti STBC can provide a full transmit diversity scheme for a system with two transmitting and $M$ receiving antennas.
The receiver decoding is provided by determining channel estimation matrix.
Based on this concept, a cooperative MIMO system is implemented using the USRP N210 and GNU radio companion software.
We have implemented a BPSK modulated Alamouti-STC system by developing Out-of Tree Modules since the GNU-radio software does not have any default Alamouti encoding or decoding blocks.
The packet structure used for Alamouti scheme implementation is shown in Fig. \ref{Alapacket}.
The Encoder block possesses a set of pilot symbols which is a random number of 1's and 0's of a fixed length which is used to calculate the CSI at the receiver based upon which the transmitted data can be decoded.
Also, when a pilot is being transmitted from acoustic transmitter 1, the transmitter 2 remains silent.
The same operation happens when transmissions occur from the second antenna. 
This is done so as to improve the efficiency in estimating the channel at the receiver.
The BPSK modulated data stream which acts as the input to the Alamouti encoder is then encoded such that its output is transmitted in different time slots. 
The encoded streams are further interpolated and pulse-shaped using a Root Raised Cosine (RRC) filter.
Then the signal streams become band-limit and are being transmitted across TX1 and TX2 respectively.
These individual data streams propagate through the complex underwater channel and are received at the base station. 
The band limited signals are then down sampled and pulse shaped using a matched filter. 
The decoding block initially estimates the channel that the signals have propagated through based on which the data stream is decoded.

The Alamouti encoder consists STBC Encoder as well as pilot signals (Random number generator) that are concatenated to the data streams and act as channel coefficients. 
The frame that is transmitted from each transmitter. 
It indicates that when TX 1 is transmitting its pilot signal, TX 2 remains silent and its the same the other way around. 
In this way, different pilot signals are transmitted across the different transmitters at the same instant of time and channel estimation at the receiver with respect to the different transmitters are individually calculated.
For the decoding procedure, initially at the decoder, the state of the receiving end is set to accept the 1st pilot sequence based on which channel estimations are summed up over the entire length of the Symbol. 
When Pilot 1 is received the state is changed to Pilot 2 and the same procedure is carried out. 
Hence the mean of the entire pilot symbols is calculated and presented as the estimated channel to further decode the data streams. 
To improve the decoding accuracy, a maximum likelihood detector is used. 
The decoded stream is then demodulated to obtain the original stream of data.
The matched filter used here is obtained by correlating a known delayed signal with an unknown signal to detect the presence of the template in the unknown signal. 
It is an optimal linear filter used for maximizing the SNR.

\begin{figure}
    \centering
    \includegraphics[width=3in]{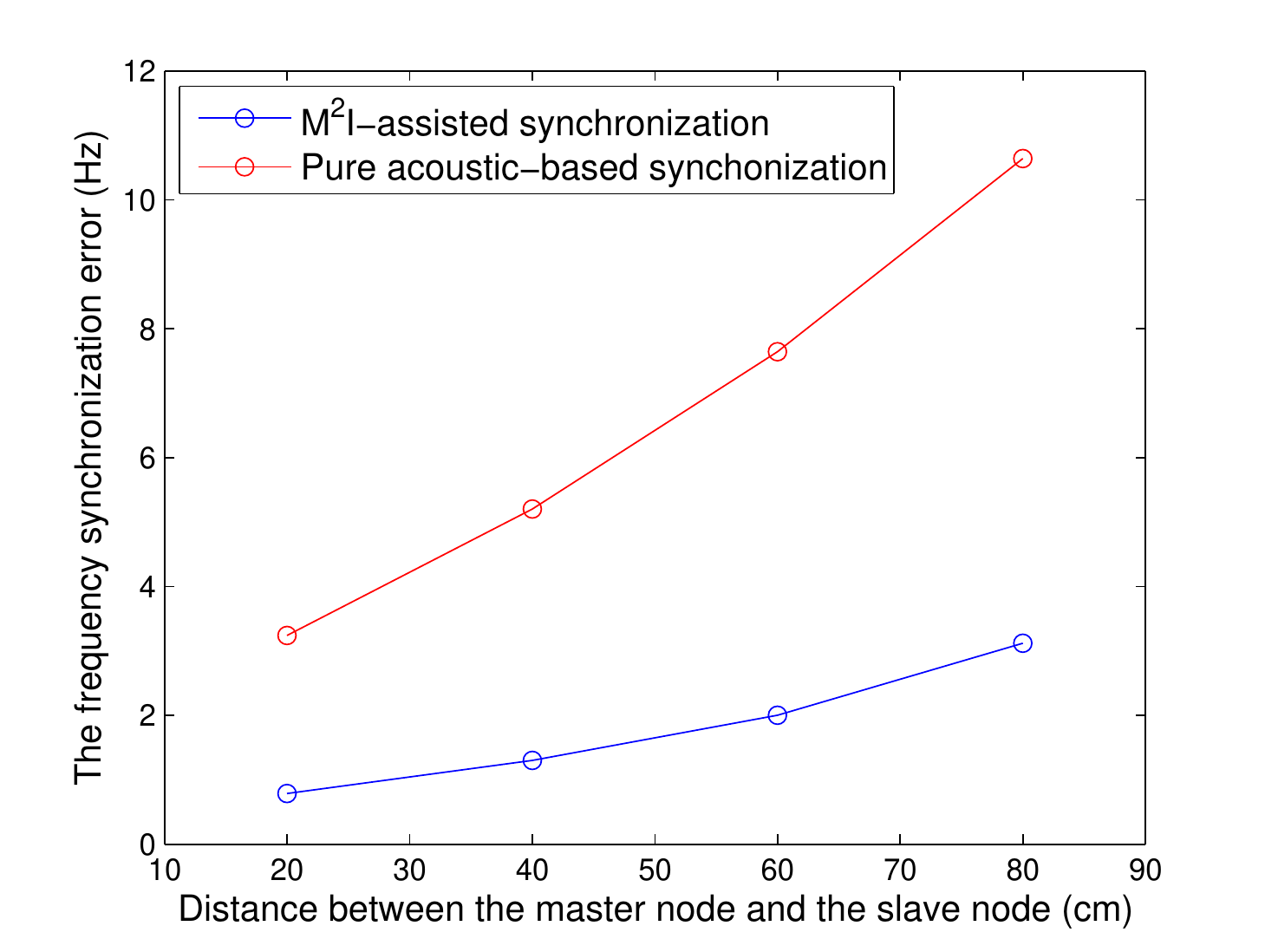}
    \caption{Carrier frequency synchronization error comparison.}
    \label{cfo}
\end{figure}

\begin{figure}
    \centering
    \includegraphics[width=3in]{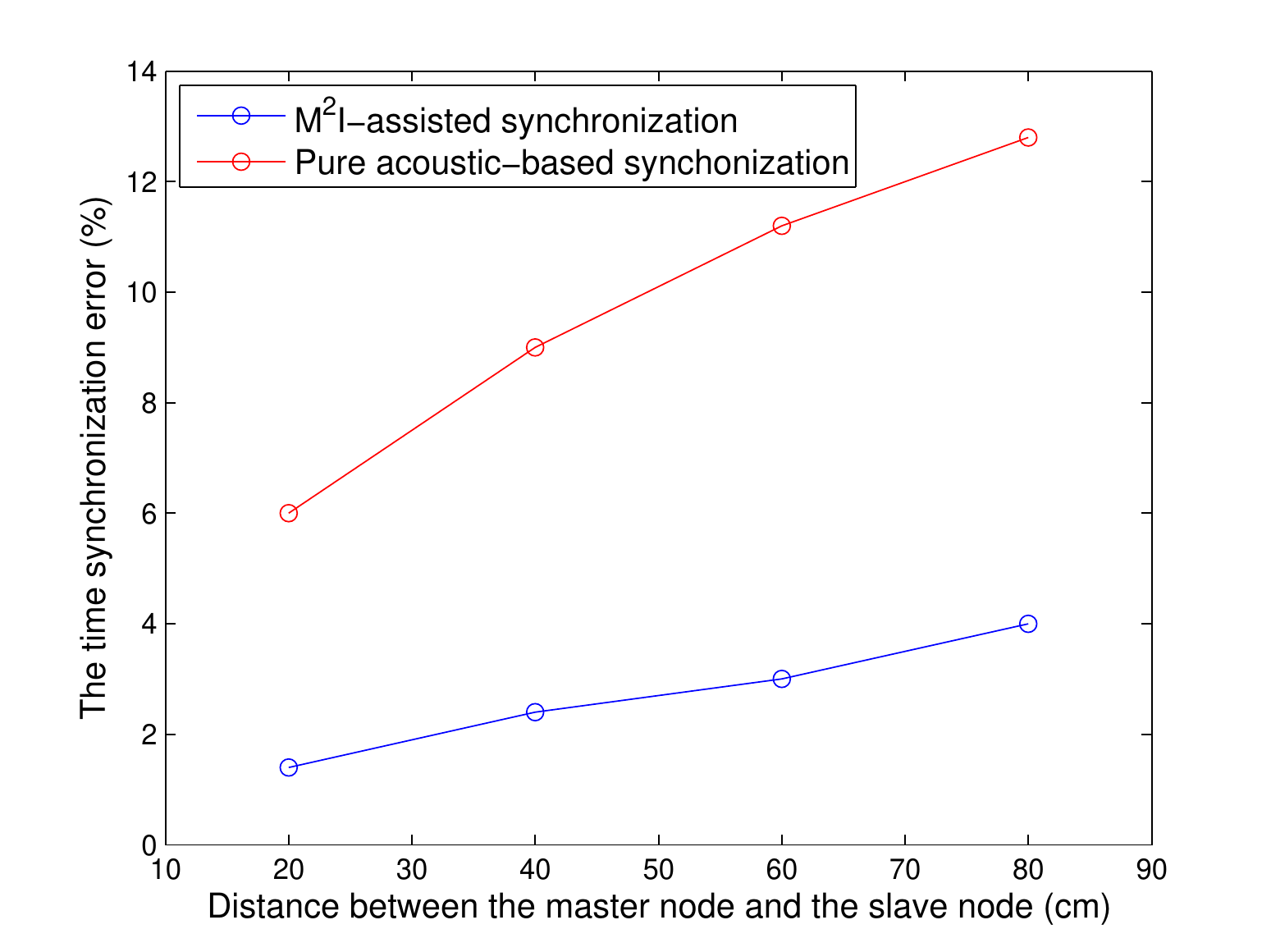}
    \caption{Time synchronization error comparison.}
    \label{to}
\end{figure}

\begin{figure}
    \centering
    \includegraphics[width=3in]{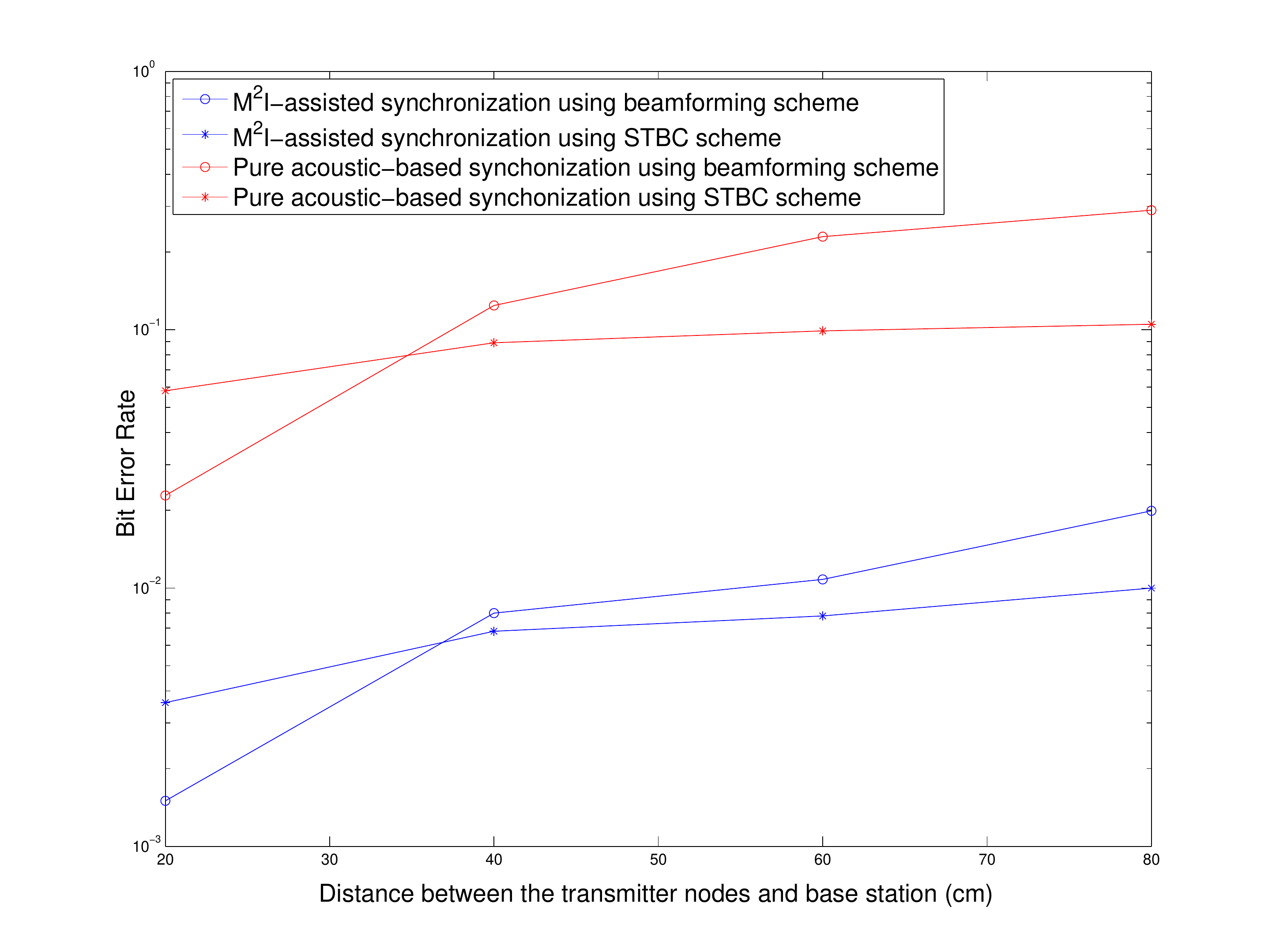}
    \caption{BER Performance using beamforming scheme and STBC scheme, two transmitters are fixed while changing the position of receiver.}
    \label{ber1}
\end{figure}

\begin{figure}
    \centering
    \includegraphics[width=3in]{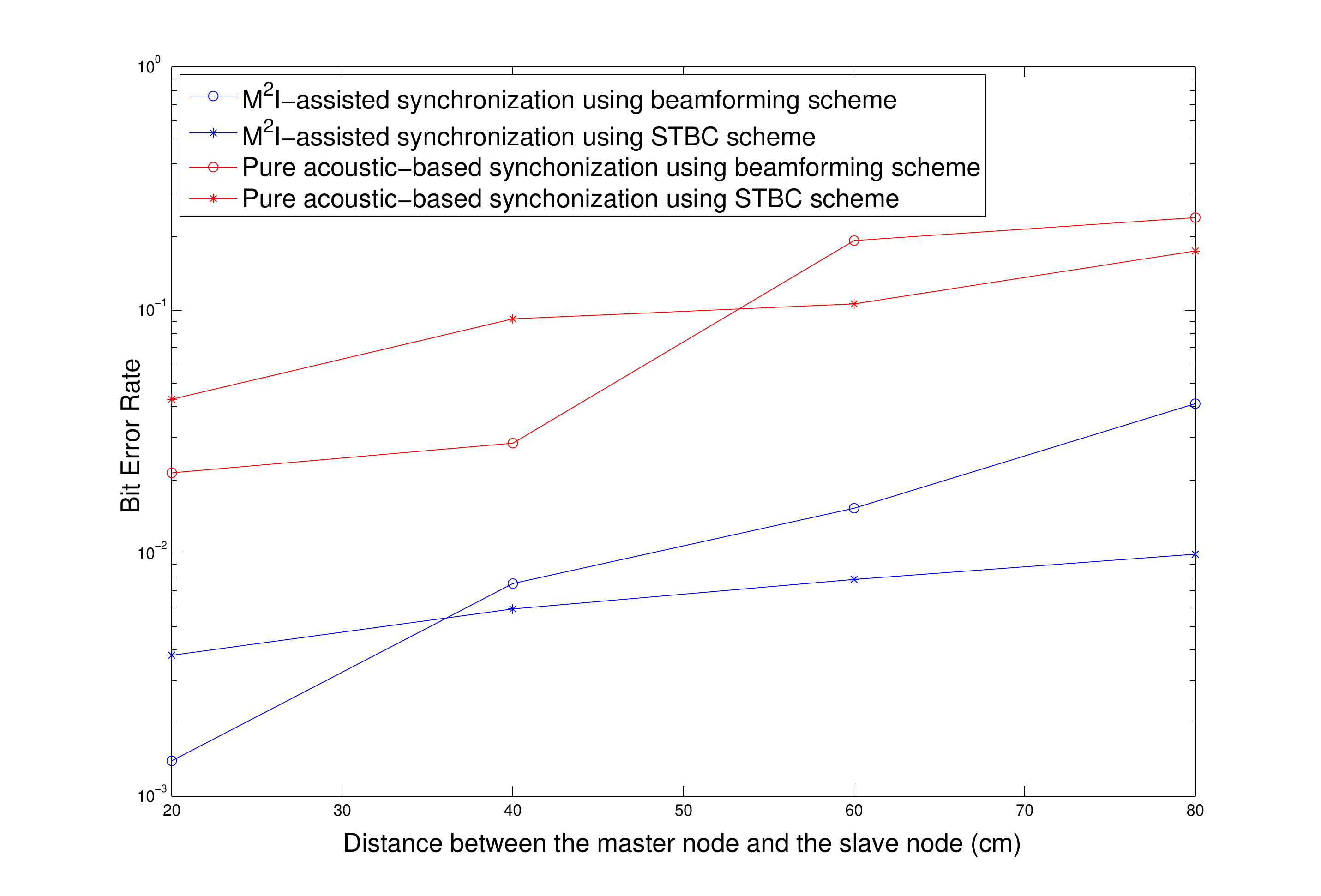}
    \caption{BER Performance using beamforming scheme and STBC scheme, two transmitters are moving while keeping the position of receiver in the middle.}
    \label{ber2}
\end{figure}

\subsection{Experimental Performance Analysis}
Experiment results for frequency and time synchronization error using pure acoustic-basic and M\(^2\)I assisted synchronization scheme are seen in Fig. \ref{cfo} and \ref{to} respectively. 
For this experiment of synchronization, data are taken by changing the distance between the antennas in slave nodes from 20 cm to 80 cm, while keep the master node in the middle.
Both acoustic antennas and  M\(^2\)I antennas are used for synchronization in order to compare their synchronization performance.
It should be noticed that the distance between each slave node is much lager than half the wavelength (0.0075m) of acoustic wave to ensure channel independence.
We can see that the Performance of the  M\(^2\)I-assisted synchronization technique  out-compete the performance of pure acoustic synchronization even when the distance between the slave nodes is large, which means our proposed M\(^2\)I-assisted synchronization scheme can reduce the frequency and time synchronization error dramatically in practical MIMO acoustic communication system.

Then we evaluate the BER performance of both Maximum SNR beamforming scheme and Alamouti STBC scheme in the experiments.
For the  Maximum SNR beamforming technique, two transmit antennas and one receive antenna are used.
The master-slave synchronization is done using acoustic antennas and M\(^2\)I antennas mentioned above.
Following the synchronization, beamforming is performed using the acoustic antennas.
The Alamouti STBC experiments are also performed using a similar experimental setup. 
In this case, there is no need for feedback from the BS to do the channel estimation, but the data is encoded using the Alamouti scheme. 
The data packet is coded using MATLAB and then given to the transmitters from GNU Radio. 

In Fig. \ref{ber1}, it shows the BER performance comparison using beamforming scheme and Alamouti STBC scheme.
The data in Fig. \ref{ber1} are taken by changing the distance between the transmitters and receiver from 20 cm to 80 cm and keeping the distance between the transmitters fixed at 30 cm.
This scenario keeps the transmitter nodes fixed and change the distance between transmitter nodes and base station, which can evaluate the BER performance between transmitters and base station.
Fig. \ref{ber2} also shows the BER performance comparison using two different schemes, however, the data in Fig. \ref{ber2} are taken by changing the distance between the transmitters and keeping the receiver approximately in the middle of the two transmitters.
In this scenario, the master node is fixed while the distance between the slave nodes is changing.
In this way, we can evaluate the BER performance between master node and slave node.

From the results shown in Fig. \ref{ber1} and Fig. \ref{ber2}, we can see that in two different scenarios and using different MIMO implementation schemes, the system using the M\(^2\)I-assisted synchronization can always achieve better overall BER performance (one order magnitude) than the system using pure acoustic synchronization.
These results demonstrate the effectiveness of our proposed hybrid  M\(^2\)I and acoustic MIMO system.
And it should be noticed that no matter M\(^2\)I-assisted or pure synchronization is used, the Alamouti STBC scheme can provide much flat BER performance than beamforming scheme.
The reason is that when using the beamforming scheme, the channel estimation process is needed, so the BER performance is largely depended on how well we estimate the channel.
When the distance between two nodes becomes far, the precise channel estimation is hard to achieve so that the BER performance deteriorates and is no more better than using Alamouti STBC scheme.
Thus, the Maximum SNR beamforming scheme is suitable in good channel condition while the Alamouti STBC scheme fits universal worse channel scenario.
This finding gives us a hint about which MIMO scheme should be implemented for the  hybrid MIMO system in real environment.

\section{Conclusion}
In this paper, we propose an underwater cooperative/distributed MIMO communication mechanism that is based on a novel hybrid $\text{M}^2\text{I}$-acoustic technique, for underwater robot swarms (URSs) as well as other underwater IoTs. The proposed cooperative MIMO system consists of two modules, i.e., the $\text{M}^2\text{I}$-assisted synchronization module and the acoustic MIMO module. The performance of the $\text{M}^2\text{I}$ module is rigorously evaluated by analyzing the synchronization accuracy of the distributed transmitters and calculating the time and frequency errors. Based on the synchronization accuracy, the communication performance of the acoustic MIMO module is evaluated under both beamforming and space-time coding schemes. Compared with the pure acoustic underwater communications, the hybrid $\text{M}^2\text{I}$-acoustic technique can truly realize the long-awaited distributed MIMO communications for small or mobile underwater devices, thanks to the much smaller propagation delay and the larger bandwidth of the MI techniques.
The performance of the proposed hybrid system, including the SNR, BER, effective communication time, and the upper bound of the throughput, are evaluated through numerical analysis, and more importantly, through real-world experiments based on a software-defined testbed that is built in-house.

\ifCLASSOPTIONcaptionsoff
  \newpage
\fi

%









\begin{thebibliography}{1}

\bibliographystyle{IEEEtran}

\bibliography{IEEEabrv,../bib/paper}

\linespread{0.92}


\bibitem{ourpaper}
S. Desai, V.D. Sudev, X. Tan, P. Wang and Z. Sun, ``Enabling Underwater Acoustic Cooperative MIMO Systems by Metamaterial-Enhanced Magnetic Induction'' in \emph{Proc. of IEEE Wireless Communications and Networking Conference (WCNC)}, Marrakech, Morocco, April 2019.

\bibitem{Akyildiz_softwater_2016}
Ian F. Akyildiz, Pu Wang, and Shih-Chun Lin, ``ASoftWater: Software-defined networking for next-generation underwater communication systems'', \emph{Ad Hoc Networks, Elsevier}, August 2016.

\bibitem{MIT_underwater_backscatter_sigcom_2019}
Junsu Jang and Fadel Adib, ``Underwater backscatter networking'', in \emph{Proc. of the ACM Special Interest Group on Data Communication (SIGCOMM '19)}, pp. 187-199, New York, NY, USA, 2019.


\bibitem{Xu_Underwater_Applications_2014}
Guobao Xu, Weiming Shen, and Xianbin Wang, ``Applications of Wireless Sensor Networks in Marine Environment Monitoring: A Survey'', \emph{Sensors Journal}, pp. 16932-16954, September 2014.

\bibitem{MariCarmen_UIoT_survay_2012}
Mari Carmen Domingo ``An overview of the internet of underwater things'', \emph{Journal of Network and Computer Applications}, Vol. 35, No. 6, pp. 1879-1890, September 2014.

\bibitem{robot1}
N. E. Leonard, D. A. Paley, R. E. Davis, D. M. Fratantoni, F. Lekien, and F. Zhang, ``Coordinated Control of An Underwater Glider Fleet in An Adaptive Ocean Sampling Field Experiment in Monterey,'' \emph{Journal of Field Robotics}, vol. 27, No. 6, pp. 718-740, September 2010.

\bibitem{robot2}
D. Paley, Fumin Zhang, and N. E. Leonard, ``Coordinated Control for Ocena Sampling: The glider Coordinated Control System,'' \emph{IEEE Transactions on Control System Technology}, vol. 16, No. 4, pp. 735-744, July 2008.

\bibitem{robot3}
S. Liu, J. Sun, J. Yu, A. Zhang, and F. Zhang, “Sampling Optimization
for Networked Underwater Gliders,” in \emph{Proc. of IEEE/MTS Oceans 2016}, Shanghai, China, 2016.

\bibitem{int1}
Jules S. Jaffe, Peter J. S. Franks, Paul L. D. Roberts, Diba Mirza, Curt Schurgers, Ryan Kastner and Adrien Boch, ``A swarm of autonomous miniature underwater robot drifters for exploring submesoscale ocean dynamics'', \emph{Nature Communications}, vol. 8, Article number: 14189, 2017.

\bibitem{int2}
Michael Bodi, Christoph Möslinger, Ronald Thenius, Thomas Schmickl, ``BEECLUST used for exploration tasks in Autonomous Underwater Vehicles'', in \emph{Proc. of 8th Vienna International Conference on Mathematical Modelling}, Vienna, Austria, February 2015.

\bibitem{int4}
Erik Sollesnes, Ole Martin Brokstad, Rolf Klæboe, Bendik Vågen, Alfredo Carella, Alex Alcocer, Artur Piotr Zolich, Tor Arne Johansen, `Towards autonomous ocean observing systems using Miniature Underwater Gliders with UAV deployment and recovery capabilities'', in \emph {Proc. of IEEE OES Autonomous Underwater Vehicle Symposium}, arXiv 1902.03112, 2019.

\bibitem{ursreview}
B. T. Champion and M. A. Joordens, “Underwater swarm robotics review,” in \emph {Proc. of 2015 10th System of Systems Engineering Conference (SoSE)}, pp. 111–116, 2015.

\bibitem{MIMO3}
J. Zhang and Y. R. Zheng, ``Frequency-Domain Turbo Equalization with Soft Successive Interference Cancellation for Single Carrier MIMO Underwater Acoustic Communications,'' \emph{IEEE Transactions on Wireless Communications}, vol. 10, No. 9, pp. 2872-2882, September 2011.

\bibitem{int3}
Ian F. Akyildiz, Pu Wang, Zhi Sun, ``Realizing underwater communication through magnetic induction'', \emph{IEEE Communications Magazine}, vol. 53, Issue 11, pp. 42 - 48, November 2015. 

\bibitem{meta}
H. Guo, Z. Sun, J. Sun, and N. M. Lichinitser, `` M\(^2\)I: Channel Modeling for Metamaterial-Enhanced Magnetic Induction Communications,'' \emph{IEEE Transactions on Antennas and Propagation}, vol. 63, No. 11, pp. 5072-5087, November 2015.

\bibitem{shallowUAC}
P. Bouvet and A. Loussert, ``MIMO underwater acoustic communications over shallow water channel'', \emph{Underwater Acoustics Lab}, ISEN Brest 2011. 

\bibitem{macmimo}
L. Kuo and T. Melodia, ``Distributed Medium Access Control Strategies for MIMO Underwater Acoustic Networking,'' \emph{IEEE Transactions on Wireless Communications}, vol. 10, no. 8, pp. 2523-2533, August 2011.

\bibitem{decisionmimo}
J. Tao, Y. R. Zheng, C. Xiao and T. C. Yang, ``Robust MIMO Underwater Acoustic Communications Using Turbo Block Decision-Feedback Equalization,'' \emph{IEEE Journal of Oceanic Engineering}, vol. 35, no. 4, pp. 948-960, October 2010.

\bibitem{sonar}
Yan Pailhas, Yvan Petillot, Keith Brown, Bernard Mulgrew, ``Spatially Distributed MIMO Sonar Systems: Principles and Capabilities,'' \emph{IEEE Journal of Oceanic Engineering}, vol. 42, No. 3, July 2017.

\bibitem{synchronization1}
D. R. Brown and H. V. Poor, ``Time-Slotted Round-Trip Carrier Synchronization for Distributed Beamforming,'' \emph{IEEE Transactions on Signal Processing}, vol. 56, No. 11, pp. 5630-5643, November 2008.

\bibitem{synchronization2}
S. Shi, S. Zhu, X. Gu, and R. Hu, ``Extendable carrier synchronization for distributed beamforming in wireless sensor networks,'' in \emph{Proc. of 2016 International Wireless Communications and Mobile Computing Conference (IWCMC)}, Paphos, Cyprus, September 2016.

\bibitem{SDR}
H Yan, S Hanna, K Balke, R Gupta and D Cabric, ``Software Defined Radio Implementation of Carrier and Timing Synchronization for Distributed Arrays,'' \emph{IEEE Aerospace Conference}, 2019.

\bibitem{disbeam}
M. Rahman, E. Baidoo-Williams, R. Mudumbai and S. Dasgupta, ``Fully wireless implementation of distributed beamforming on a software-defined radio platform'' in \emph{Proc. of the 11th international conference on Information Processing in Sensor Networks}, pages 305-316, April 2012.

\bibitem{jmb}
H. Rahul, S. Kumar and D. Katabi, ``JMB: Scaling Wireless Capacity with User Demands'' in \emph{Proc. of ACM SIGCOMM}, Helsinki, Finland, August 2012.

\bibitem{Dmimo}
G. Sklivanitis, Y. Cao, S. N. Batalama and W. Su, ``Distributed MIMO Underwater Systems: Receiver Design and Software-Defined Testbed Implementation,'' in \emph{Proc. of 2016 IEEE Global Communications Conference (GLOBECOM)}, Washington DC, USA, 2016.

\bibitem{dopplercoop}
Kai Tu, Tolga M. Duman, John G. Proakis, Milica Stojanovic, ``Cooperative MIMO-OFDM communications: Receiver design for Doppler-distorted underwater acoustic channels,'' in \emph{Proc. of 4th Asilomar Conference on Signals, Systems and Computers}, November 2010.

\bibitem{underwaterMI}
H. Guo, Z. Sun, and P. Wang, ``Channel modeling of MI Underwater Communication Using Tri-Directional Coil Antenna,'' in \emph{Proc. of IEEE Globecom 2015}, San Diego, USA, December 2015.

\bibitem{M2I}
H. Guo and Z. Sun, ``Demo abstract: Prototyping M2I communication system for underground and underwater networks,'' \emph{2017 IEEE Conference on Computer Communications Workshops (INFOCOM WKSHPS)}, Atlanta, GA, 2017, pp. 962-963.

\bibitem{metashell}
H. Guo, Z. Sun and C. Zhou,``Practical Design and Implementation of Metamaterial-Enhanced Magnetic Induction Communication," \emph{IEEE Access}, vol. 5, pp. 17213-17229, 2017.


\bibitem{realmimo}
Ezzeldin Hamed, Hariharan Rahul, Mohammed A. Abdelghany, and Dina Katabi. "Real-time Distributed MIMO Systems" in \emph{Proc. of the 2016 ACM SIGCOMM Conference (SIGCOMM '16)}, pp. 412-425, New York, NY, USA, 2016.

\bibitem{metaextreme}
H. Guo and Z. Sun,``Full-duplex Metamaterial-enabled Magnetic Induction Networks in Extreme Environments'' in \emph{Proc. of IEEE Conference on Computer Communications (INFOCOM)}, pp. 558-566, Honolulu, HI, USA, April 2018.

\bibitem{coherent}
Y.-S. Tu and G. J. Pottie. ``Coherent cooperative transmission from multiple adjacent antennas to a distant stationary antenna through AWGN channels,'' in \emph{Proc. of IEEE 55th Vehicular Technology Conference (VTC Spring 2002)}, 2002.

\bibitem{chirpsignal}
H. Kulhandjian, T. Melodia and D. Koutsonikolas,``CDMA-Based Analog Network Coding for Underwater Acoustic Sensor Networks,''  \emph{IEEE Transactions on Wireless Communications}, vol. 14, no. 11, pp. 6495-6507, November 2015.

\bibitem{model}
M. Naderi, M. Patzold, R. Hicheri, and N. Youssef, ``A Geometry-Based Underwater Acoustic Channel Model Allowing for Sloped Ocean Bottom Conditions,'' \emph{IEEE Transactions on Wireless Communications}, vol. 16, No. 4, pp. 2394-2408, April 2017.

\bibitem{coherence}
T. S. Rappaport, ``Wireless Communications: Principles and Practive,'' \emph{New Jersey: Presntice Hall PTR, Upper Saddle River}, 2001.

\bibitem{noise}
E. Demirors, G. Sklivanitis, T. Melodia, S. Batalama, and D. Pados, ``Software-defined Underwater Acoustic Networks: Towards a High-rate Real-time Reconfigurable Modem,'' \emph{IEEE Communications Magazine}, vol. 53, No. 11, pp. 64-71, November 2015.

\bibitem{cfo}
J. Li, G. Liu, and G. B. Giannakis,``Carrier frequency offset estimation for OFDM-based WLANs," \emph{IEEE Signal Processing Letter}, vol. 8, pp. 80–82, 2001.

\bibitem{maxlike}
Y. Cao, W. Su and S. N. Batalama, ``A Novel Receiver Design and Maximum-Likelihood Detection for Distributed MIMO Systems in Presence of Distributed Frequency Offsets and Timing Offsets,'' \emph{IEEE Transactions on Signal Processing}, vol. 66, no. 23, pp. 6297-6309, December 2018.


\end{thebibliography}
\end{document}